\documentclass[a4paper,fleqn,usenatbib]{mnras}
\usepackage{mathtools}

\usepackage{breqn}
\usepackage[T1]{fontenc}
\usepackage{ae,aecompl}
\usepackage{tablefootnote}

\usepackage{graphicx}	% Including figure files
\usepackage{amsmath}	% Advanced maths commands
\usepackage{amssymb}	% Extra maths symbols
\usepackage{subfig}
\usepackage{float}
\usepackage{multirow}
\usepackage{units}
\usepackage{placeins}

\usepackage{xcolor}

\title[HFQPOs of GRS 1915$+$105]
{{Wide-band view of High Frequency QPOs of GRS 1915$+$105 in `softer' variability classes observed with \it AstroSat}}

\author[Majumder et al.]{Seshadri Majumder$^{1}$\thanks{E-mail: smajumder@iitg.ac.in},
H. Sreehari$^{2}$, 
Nafisa Aftab$^{1}$, 
Tilak Katoch$^{3}$, 
\newauthor Santabrata Das$^{1}$\thanks{E-mail: sbdas@iitg.ac.in}, Anuj Nandi$^{4}$\thanks{E-mail: anuj@ursc.gov.in} \\
$^{1}$Department of Physics, Indian Institute of Technology Guwahati, Guwahati, 781039, India.\\
$^{2}$Indian Institute of Astrophysics, Bangalore, 560034, India. \\
$^{3}$DAA, Tata Institute of Fundamental Research, Colaba, Mumbai, 400005, India.\\
$^{4}$Space Astronomy Group, ISITE Campus, U. R. Rao Satellite Centre, Outer
Ring Road, Marathahalli, Bangalore, 560037, India.
}

\date{Accepted XXX. Received YYY; in original form ZZZ}

\pubyear{2021}

\begin{document}
\label{firstpage}
\pagerange{\pageref{firstpage}--\pageref{lastpage}}
\maketitle

%\end{document}
%%%%%%%%%%%%%%%%%%%%%%%%%%%%%%%%%%%%% ABSTRACT %%%%%%%%%%%%%%%%%%%%%%%%%%%%%%%%%%%%%%%%%%%%%%%%%%%%%%
\begin{abstract}
		
We present a comprehensive temporal and spectral analysis of the `softer' variability classes ($i.e.$, $\theta$, $\beta$, $\delta$, $\rho$, $\kappa$, $\omega$ and $\gamma$) of the source GRS 1915+105 observed by {\it AstroSat} during $2016-2021$ campaign. Wide-band ($3-60$ keV) timing studies reveal the detection of High Frequency Quasi-periodic Oscillations (HFQPOs) with frequency of $68.14-72.32$ Hz, significance of $2.75-11 \sigma$, and rms amplitude of $1.48-2.66\%$ in $\delta$, $\kappa$, $\omega$ and $\gamma$ variability classes. Energy dependent power spectra impart that HFQPOs are detected only in $6-25$ keV energy band and rms amplitude is found to increase ($1-8\%$) with energy. The dynamical power spectra of $\kappa$ and $\omega$ classes demonstrate that HFQPOs seem to be correlated with high count rates. We observe that wide-band ($0.7-50$~keV) energy spectra can be described by the thermal Comptonization component (\texttt{nthComp}) with photon index ($\Gamma_{\rm nth}$) of $1.83-2.89$ along with an additional steep ($\Gamma_{\rm PL}\sim3$) \texttt{powerlaw} component. The electron temperature ($kT_e$) of $1.82 -3.66$ keV and optical depth ($\tau$) of $2-14$ indicate the presence of a cool and optically thick corona. In addition, \texttt{nthComp} components ($1.97 \lesssim \Gamma_{\rm nth} \lesssim 2.44$, $1.06 \times 10^{-8}  \lesssim F_{\rm nth} ~({\rm erg} {\rm ~cm}^{-2} {\rm ~s}^{-1}) \lesssim 4.46\times 10^{-8}$) are found to dominate in presence of HFQPOs. Overall, these findings infer that HFQPOs are possibly resulted due to the modulation of the `Comptonizing corona'. Further, we find that the bolometric luminosity ($0.3-100$ keV) of the source lies within the sub-Eddington ($3-34\%$ $L_{\rm Edd}$) regime. Finally, we discuss and compare the obtained results in the context of existing models on HFQPOs.

\end{abstract}
%%%%%%%%%%%%%%%%%%%%%%%%%%%%%%%%%%%%%%%%%%%%%%%%%%%%%%%%%


%\clearpage
%%%%%%%%%%%%%%%%%%%%%%%%%%%%%%%%%%%%%%%%%Introduction%%%%%%%%%%%%%%%%%%%%%%%%%%%%%%%%%
\section{Introduction}

The black hole X-ray binaries (BH-XRBs) occasionally exhibit High Frequency Quasi-periodic Oscillation (HFQPO) features that are potentially viable to probe the effect of strong gravity in the vicinity of the compact objects. The signature of QPOs is observed as a narrow feature with excess power in the power density spectrum \citep{VanderKlis1989}. In general, QPO frequencies are classified in two different categories in BH-XRB systems, namely (a) Low-frequency QPO (LFQPO) with centroid frequency $\nu_{\rm QPO} < 40$ Hz, and (b) High-frequency QPO (HFQPO) with $\nu_{\rm QPO}$ exceeding $40$ Hz \citep{Remillard-McClintock2006}. LFQPOs are common in BH-XRB systems, whereas HFQPOs are detected in few BH-XRBs observed with {\it RXTE}\footnote{https://heasarc.gsfc.nasa.gov/docs/xte/XTE.html}, such as, GRS 1915+105 ($65-69$ Hz, \citealt{Morgan-etal1997}, \citealt{Belloni-etal2013}), GRO J1655$-$40 ($300$ and $450$ Hz, \citealt{Remillard-etal1999}, \citealt{Strohmayer2001a}, \citealt{Remillard-etal2002}), XTE J1550$-$564 ($102-284$ Hz, \citealt{Homan-etal2001}; $188$ Hz and $249-276$ Hz, \citealt{Miller-etal2001}), H 1743$-$322 ($160$ and $240$ Hz, \citealt{Homan-etal2005}; $166$ Hz, $239$ Hz and $242$ Hz, \citealt{Remillard-etal2006}), XTE J1650$-$500 ($50$ Hz and $250$ Hz, \citealt{Homan-etal2003}), 4U 1630$-$47 ($100-300$ Hz, \citealt{Klein-Wolt-etal2004}), XTE J1859+226 ($150$ Hz and $187$ Hz, \citealt{Cui2000}) and IGR J17091$-$3624 ($66$ Hz and $164$ Hz, \citealt{Altamirano-Belloni2012}), respectively. However, the detection of HFQPOs in XTE J1650$-$500, 4U 1630$-$47, and XTE J1859+226 remain inconclusive due to their broad features with lesser significance \citep{Belloni-etal2012}.

The HFQPOs are detected by {\it RXTE} in high flux observations with intermediate hardness ratios. However, it is intriguing that not all observations with high flux do show this feature \citep{Belloni-etal2012}. Typically, HFQPOs are observed with either one or two peaks in power spectra and the corresponding centroid frequencies are found to vary with time \citep{Belloni-etal2014}. In some instances, the simultaneous observations of HFQPOs of $\sim 3:2$ frequency ratio are reported in GRO J1655$-$40 and H1743$-$322 \citep{Strohmayer2001a,Remillard-etal2002}, which are possibly yielded due to the resonance between two epicyclic oscillation modes \citep{Abramowicz-etal2001}. Further, the fractional variabilities ($i.e.$, rms amplitudes) of HFQPOs, detected in various sources, are found to increase with energy \citep{Miller-etal2001}. In particular, for GRS 1915+105, the percentage rms associated with $67$ Hz increases from $1.5\%$ (at $\sim 5$ keV) to $6\%$ (at $\sim 20$ keV) \citep{Morgan-etal1997}.  
In addition, \citealt{Homan-etal2001} measured time lags for $\sim 282$ Hz HFQPO and found either zero or negative time lags (soft photons lag hard photons) for XTE J$1550-564$ source. Also, hard phase lags (hard photons lag soft photons) were observed for $67$ Hz HFQPO in GRS 1915$+$105 including other BH-XRBs as well \citep{Cui-etal1999,Mendez-etal2013}. On the other hand, a negative phase lag was observed for $35$ Hz feature, which appeared simultaneously with $67$ Hz HFQPO in GRS $1915+105$. Interestingly, the magnitude of these soft and hard lags are found to increase with energy \citep{Mendez-etal2013}.

Needless to mention that the thermal and non-thermal spectral components of BH-XRB spectra reveal the characteristics of the underlying emission processes and the geometry of the accretion disc. In general, the thermal emissions are mostly originated from the different radii of the multi-temperature accretion disc \citep{Shakura-etal1973}, whereas the high energy non-thermal emissions are emanated due to inverse-Compton scattering of seed blackbody photons reprocessed at the `hot' corona surrounding the inner part of the accretion disc (\citealt[and references therein]{Sunyaev-etal1980,Tanaka-etal1995,Chakrabarti-etal1995,Mandal-Chakrabarti2005,Iyer-etal2015}). Alternative prescriptions of the jet based corona model are also widely discussed in the literature 	 \citep{Beloborodov1999,Fender-etal1999,Wang-etal2021,Lucchini-etal2021} including different coronal geometries \citep{Haardt-etal1993,Markoff-etal2005,Nowak-etal2011,Poutanen-etal2018}. Often, the presence of HFQPOs is found to be prominent largely in the softer states dominated by the disc emission \citep{McClintock-etal2006}. Therefore, it is important to carry out the wide-band spectral modeling in the presence of HFQPO features.

GRS 1915$+$105, known as microquasar \citep{Mirabel-Rodriguez1994}, is a very bright BH-XRB system which was discovered in $1992$ with {\it GRANAT} mission \citep{Castro-Tirado-etal1992}. The source possibly harbors a fast spinning Kerr black hole \citep{McClintock-etal2006} with spin $>0.98$ measured by indirect means \citep[and references therein]{Sreehari-etal2020}. The mass and distance of the black hole are constrained as $12.4_{-1.8}^{+2.0}$ M$_{\odot}$ and 8.6 kpc, respectively \citep{Reid-etal2014}. Interestingly, GRS $1915+105$ shows different types of structured variabilities in its light curves with time scale of seconds to minutes, and these are identified into $14$ distinct classes (\citealt{Belloni-etal2000}, \citealt{Klein-wolt-etal2002}, \citealt{Hannikainen-etal2005}). Meanwhile, \textit{RXTE} extensively observed both LFQPOs (\citealt{Nandi-etal2001,Vadawale-etal2001,Ratti-etal2012} and references therein) and HFQPOs (\citealt{Morgan-etal1997}; \citealt{Strohmayer2001a}; \citealt{Belloni-etal2006};  \citealt{Belloni-etal2013};  \citealt{Mendez-etal2013}) in this source. \cite{Morgan-etal1997} first detected $65-67$ Hz HFQPO in GRS $1915+105$ observed with \textit{RXTE}. \cite{Belloni-etal2006} reported the detection of HFQPO with frequency $170$ Hz in $\theta$ class, whereas $63 - 71$ Hz HFQPO is observed in $\kappa$, $\gamma$, $\mu$, $\delta$, $\omega$, $\rho$, and $\nu$ classes as well \citep{Belloni-etal2013}. Simultaneous detection of $34$ and $41$ Hz features with the fundamental HFQPO at $\sim 68$ Hz was also reported with \textit{RXTE} (\citealt{Strohmayer2001b}; \citealt{belloni-etal2013b}).

Recently, using \textit{AstroSat} observations, \cite{Belloni-etal2019} and \cite{Sreehari-etal2020} observed HFQPO of frequencies $67.4-72.3$ Hz and $67.96-70.62$ Hz in GRS 1915$+$105, respectively. \cite{Belloni-etal2019} studied the temporal properties of GRS 1915+105 considering only two variability classes from 2017 observations. They found a direct correlation between the centroid frequency of HFQPOs and hardness, including positive phase lags which were found to increase with energy and decrease with hardness. However, they did not investigate the spectral characteristics of the source. Further, \cite{Sreehari-etal2020} observed the gradual decrease of the strength of the HFQPO features in $\delta$ class that eventually disappear with the increase of the count rate and the decrease of hardness ratio.  They also infer that HFQPOs are present in the $6-25$ keV energy range and ascertain that the HFQPOs in GRS $1915+105$ seem to be yielded due to an oscillating Comptonized `compact' corona surrounding the central source.

In this paper, for the first time to the best of our knowledge, we carry out in-depth analysis and modeling of wide-band \textit{AstroSat} observations of eight variability classes ($\theta$, $\beta$, $\delta$, $\rho$, $\kappa$, $\omega$, $\gamma$ and $\chi$) of GRS 1915$+$105 during $2016-2021$ to study the HFQPO features. While doing so, we examine the color-color diagram by defining the soft color (HR1, ratio of count rates in $6-15$ keV to $3-6$ keV) and hard color (HR2, ratio of count rates in $15-60$ keV to $3-6$ keV). Adopting the selection criteria for the `softer' variability classes as $0.02 \lesssim {\rm HR2} \lesssim 0.11$ and $0.61 \lesssim {\rm HR1} \lesssim 0.90$, and `harder' variability class with ${\rm HR2} > 0.11$ and ${\rm HR1} \gtrsim 0.7$, we find seven `softer' variability classes, namely $\delta$, $\rho$, $\kappa$, $\omega$, $\gamma$, $\beta$ and $\theta$, respectively. Subsequently, we examine the light curves and study the energy dependent HFQPO features, percentage rms variabilities ($rms\%$), and dynamic power spectra using {\it LAXPC} observations. We find HFQPO features in $\delta$, $\kappa$, $\omega$, and $\gamma$ variability classes, whereas no such HFQPO signatures are seen in $\theta$, $\beta$, $\rho$ and $\chi$ classes. We model the wide-band ($0.7-50$ keV) energy spectra by combining {\it SXT} and {\it LAXPC} data to understand the characteristics of the emission processes. Finally, we attempt to correlate the temporal and spectral parameters to explain the underlying mechanism responsible for the generation of HFQPO phenomena in the `softer' variability classes of GRS 1915 + 105 observed with \textit{AstroSat}.

The paper is organized as follows. In \S 2, we discuss the observations and data reduction procedures of the {\it SXT} and {\it LAXPC} instruments.  In \S 3, we present the characteristics of different variability classes ($i.e.$, $\theta$, $\beta$, $\delta$, $\rho$, $\kappa$, $\omega$, $\gamma$ and $\chi$) and discuss the results of both static and dynamic analyses of the power density spectra. Results from wide-band spectral analysis with and without HFQPO features are presented in \S 4. We discuss the results from spectro-temporal correlation in \S 5. In \S 6, we present a discussion based on the results from temporal and spectral studies in the context of the existing models of HFQPOs for BH-XRBs. Finally, we conclude in \S 7.

%%%%%%%%%%%%%%%%%%%%Table-1 %%%%%%%%%%%%%%%%%%%%%%%%%%%%%%%%%%%%%%%%%%%%%%

\begin{table*}
	\caption{Observation details of the source GRS 1915$+$105 observed by {\it AstroSat} during 2016 to 2021 in seven `softer' and one `harder' variability classes. In the table, ObsID along with MJD, Orbit number and exposure time are mentioned. The detected ($\rm r_{det}$) and incident ($\rm r_{in}$) count rate of the {\it LAXPC} detector along with hardness ratios are also tabulated. MJD 57451 corresponds to $4^{\rm th}$ March, 2016. See text for details.}
	\resizebox{1.0\textwidth}{!}{%
		\begin{tabular}{c c c c c c c c c c c c c c c c c c c}
			\hline\hline
			\multirow{2}{*}{ObsID} & \multirow{2}{*}{MJD} & \multirow{2}{*}{Orbit} & \multirow{2}{*}{Effective} & \multirow{2}{*}{$\rm r_{det}$} & \multirow{2}{*}{$\rm r_{in}$} & \multirow{2}{*}{HR1} & \multirow{2}{*}{HR2} & \multirow{2}{*}{Variability} & \multirow{2}{*}{HFQPO} & \\ \\
			& & & Exposure (s)& (cts/s)& (cts/s)& (B/A)$^*$& (C/A)$^*$& Class& & \\
			\hline
			
			T01\_030T01\_9000000358 & 57451.89 & 2351& 3459 & 7252 & 8573 & 0.68 & 0.07 & $\theta$& No &\\
			
			& 57452.82 & 2365 & 3148 & 4376 & 4825 & 0.69 & 0.09 & $\chi$ & No &\\
			
			& 57453.35 & 2373 & 1099 & 8915 & 10997 & 0.69 & 0.09 & $\theta$ & No & \\
			
			G05\_214T01\_9000000428 & 57504.02 & 3124 & 2533 & 7308 & 8651 & 0.76 & 0.04 & $\omega$& Yes &\\
			
			G05\_189T01\_9000000492 & 57552.56& 3841& 3027& 7965& 9588& 0.75& 0.03& $\delta$& No&\\
			
			& 57553.88 & 3860& 2381& 6744& 7872& 0.87& 0.06& $\delta$& Yes&\\
			
			G06\_033T01\_9000000760 & 57689.10 & 5862 & 2674 & 5359 & 6047 & 0.62 & 0.04 & $\beta$ & No & \\
			
			G06\_033T01\_9000000792 & 57705.22 & 6102& 3633& 6712& 7829& 0.68& 0.02& $\delta$& No&\\
			
			G07\_028T01\_9000001232 & 57891.88 & 8863& 3228& 1431& 1476& 0.73& 0.11& $\rho \textprime$ & No&\\
			
			G07\_046T01\_9000001236 & 57892.74 & 8876& 3627& 1292& 1328& 0.61& 0.09& $\rho$& No&\\
			
			G07\_028T01\_9000001370 & 57943.69 & 9629& 977& 2814& 2993& 0.76& 0.04& $\kappa$& No& \\
			
			& 57943.69 & 9633& 3442& 2701& 2866& 0.75& 0.04& $\kappa$& Yes&\\
			
			G07\_046T01\_9000001374 & 57946.10 & 9666& 1093& 3115& 3335& 0.79& 0.04& $\kappa$& Yes&\\
			
			& 57946.34 & 9670& 2735& 3160& 3387& 0.79& 0.04& $\kappa$& Yes&\\
			
			G07\_028T01\_9000001406 & 57961.39 & 9891& 1323& 3642& 3947& 0.86& 0.07& $\kappa$& No& \\
			
			& 57961.39 & 9894& 2451& 4415& 4872& 0.87& 0.05& $\kappa$& Yes&\\
			
			G07\_046T01\_9000001408 & 57961.59 & 9895& 3590& 4726& 5254& 0.87& 0.05& $\kappa$& Yes&\\
			
			G07\_028T01\_9000001500 & 57995.30 & 10394& 3036& 5719& 6511& 0.90& 0.06& $\omega$& Yes&\\
			
			G07\_046T01\_9000001506 & 57996.46 & 10411& 3632& 6160& 7088& 0.88& 0.05& $\omega$& Yes&\\
			
			G07\_046T01\_9000001534 & 58007.80 & 10579& 1729& 6968& 8180& 0.84& 0.05& $\omega$& Yes&\\
			
			& 58008.08 & 10583& 1898& 7392& 8769& 0.88& 0.05& $\gamma$& Yes&\\
			
			A04\_180T01\_9000001622 & 58046.36 & 11154& 2059& 7312& 8657& 0.66& 0.02& $\delta$& No&\\
			
			A04\_180T01\_9000002000 & 58209.13 & 13559 & 2632 & 1403 & 1445 & 0.87 & 0.19 & $\chi$ & No & \\
			
			A05\_173T01\_9000002812 & 58565.82 & 18839 & 3626 & 300 & 302 & 1.02 & 0.28 & $\chi$ & No & \\
			
			\hline\hline
		\end{tabular}%
	}
	\label{table:Obs_details}
	\begin{list}{}{}
		\item[$^*$]A, B and C are the count rates in $3-6$ keV, $6-15$ keV and $15-60$ keV energy ranges, respectively \cite[see][]{Sreehari-etal2020}.
	\end{list}
\end{table*}

\section{Observation and Data Reduction}
\label{s:Data-reduction}

India's first multi-wavelength space-based observatory {\it AstroSat} \citep{Agrawal2006} provides a unique opportunity to observe various astrophysical objects in the X-ray band of $0.3-100$ keV energy range. It consists of three basic X-ray instruments, namely Soft X-ray Telescope ({\it SXT}) \citep{Singh-etal2017}, Large Area X-ray Proportional Counter ({\it LAXPC}) \citep{Yadav-etal2016, Agrawal-etal2017, Antia-etal2017} and Cadmium Zinc Telluride Imager ({\it CZTI}) \citep{Vadawale-etal2016}. The source GRS 1915$+$105 was observed by {\it AstroSat} for $47$ pointed observations (termed as ObsID) during different time periods in between $2016-2021$. In this work, we examine all $47$ ObsIDs that include 38 Guaranteed Time (GT) data, 4 Announcement of Opportunity (AO) cycle data, and 5 Target of Opportunity (TOO) cycle data of {\it LAXPC} and {\it SXT} instruments. These observations exhibit all together seven `softer' ($\theta$, $\beta$, $\delta$, $\rho$, $\kappa$, $\omega$ and $\gamma$) and one `harder' ($\chi$) variability classes, respectively. In order to avoid repetition of results from identical variability classes, we consider $24$ Orbits from $17$ ObsIDs as delineated in Table \ref{table:Obs_details}.

\subsection{Soft X-ray Telescope (SXT)}

\begin{figure}
	\begin{center}
		\includegraphics[width=0.47\textwidth]{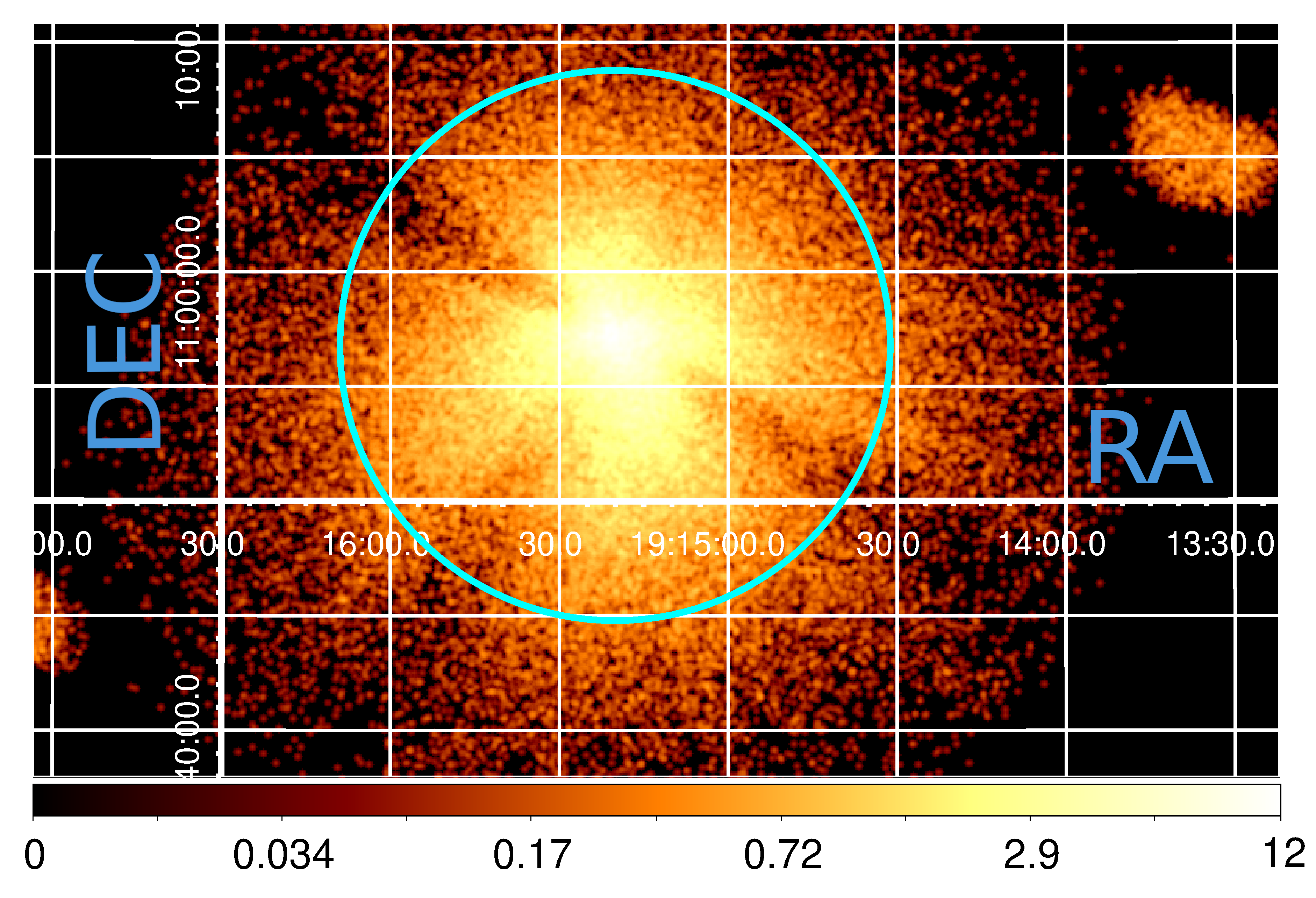}
	\end{center}

	\caption{{\it SXT} image of the source GRS 1915$+$105 obtained with the PC mode data in $\omega$ class observation on MJD 57995.30 (Orbit 10394). The cyan circle corresponds to $12$ arcmin circular region, which is considered as the source region. The colormap denotes the intensity distribution of the source. See text for details.}
	\label{fig:Source_image}
\end{figure}

{\it SXT} is a Charged Coupled Device (CCD) based X-ray imaging instrument onboard {\it AstroSat} in the energy range of $0.3-8$ keV, which operates both in Fast Window (FW) and Photon Counting (PC) modes. {\it SXT} data is analyzed following the guidelines provided by the {\it SXT} instrument team\footnote{\url{https://www.tifr.res.in/~astrosat_sxt/index.html}}. We obtain level-2 {\it SXT} data from the Indian Space Science Data Center (ISSDC)\footnote{\url{https://webapps.issdc.gov.in/astro_archive/archive/Home.jsp.}} archive. For all the observations under consideration, {\it SXT} data are available in PC mode except two observations (Orbit 3841, 3860) which are available in FW mode. The source images, light curves, and spectra are generated from level-2 cleaned event file using \texttt{XSELECT V2.4g} in \texttt{HEASOFT V6.26.1}. While examining the pile-up effect, we find that the source counts are less than $0.5$ cts$/$pixel$/$frame and less than $40$ cts/s in the central $1$ arcmin circular region of the image. Hence, we do not incorporate the pile-up correction in our analysis following the {\it AstroSat Handbook}\footnote{\url{http://www.iucaa.in/~astrosat/AstroSat_handbook.pdf.}} (see also \cite{Baby-etal2020, Katoch-etal2021}, for details). Further, we consider two scenarios of $12$ arcmin and $14$ arcmin circular regions concentric with the source coordinate for PC mode data. Extracting source counts within these two regions separately, we find that $12$ arcmin region contains up to $90\%$ of the total photon counts. Hence, in our analysis, we choose $12$ arcmin circular region as the source region (Fig. \ref{fig:Source_image}) for all the PC mode data and extract the source images, light curves, and spectra from this region. For the observations corresponding to FW mode data, we find that the source was in offset in the CCD frame \citep{Sreehari-etal2020}. Because of this, we choose the source region as $5$ arcmin circular region for these two observations. Since the timing resolution ($\sim 2.38$ s) of {\it SXT} is poor compared to {\it LAXPC}, we use {\it SXT} data only for the spectral analysis. In this work, we use the {\it SXT} instrument response file, the background spectrum file, and the ancillary response file (ARF) for both PC and FW mode data provided by the {\it SXT} instrument team\footnote{\url{https://www.tifr.res.in/~astrosat_sxt/dataanalysis.html.}}.

%%%%%%%%%%%%%%%%%%%%%%%%% Figure-1%%%%%%%%%%%%%%%%%%%%%%%%%%%%%%%%%%%%%%
%%%%%%%%%%%%%%%%%%%%%%%%%%%%%%%%%%%%%%%%%%%%%%%%%%%%%%%%%%%%%%%%%%%%%%%%%%

\subsection{Large Area X-ray Proportional Counter (LAXPC)}

{\it LAXPC} is a proportional counter consisting of three identical detectors {\it LAXPC10},  {\it LAXPC20} and  {\it LAXPC30} having combined effective area of 6000 ${\rm cm}^{2}$, which operates in $3-80$ keV energy range \citep{Yadav-etal2016, Agrawal-etal2017, Antia-etal2017}. All the three {\it LAXPC} detectors have a temporal resolution of $10 ~\mu s$ that offers rich timing analysis compared to {\it SXT}. We use {\it LAXPC} level-1 data in event analysis mode available in {\it AstroSat} public archive$^3$
for timing as well as spectral analyses. The details of {\it LAXPC} data extraction procedure and analysis methods are mentioned in \cite{Sreehari-etal2019, Sreehari-etal2020}. The software \texttt{LaxpcSoftv3.4}\footnote{\url{http://www.tifr.res.in/~astrosat_laxpc/LaxpcSoft.html}}\citep{Antia-etal2017}, released on June 14, 2021 is used to process the level-1 data to level-2 data. We extract data from the top layer of the detector and consider only single events in our analysis. Further, we choose the background models, which are generated closest to the observation dates. While doing data extraction, {\it LAXPC} instrument response files are generated following \cite{Antia-etal2017}. The software generates the Good Time Interval (GTI) of the data consisting of the observation's time information excluding the data gap due to Earth occultation and South Atlantic Anomaly (SAA). The longest continuous observation in each orbit is considered in the present analysis (see Table \ref{table:Obs_details}). Background subtracted {\it LAXPC10} and {\it LAXPC20} combined light curve of $1$ s time resolution in different energy ranges are generated by choosing the corresponding {\it LAXPC} channels using the standard routine of the software. It may be noted that we extract the source spectra from {\it LAXPC20} data only as it's gain remain stable throughout the entire observational period \citep{Antia-etal2021}.

%%%%%%%%%%%%%%%%%%% Table-2 %%%%%%%%%%%%%%%%%%%%%%%%%%%%%%%%%%%%%%%%%%%

\begin{table*}
	\centering
	\caption{Details of the best fitted PDS parameters from {\it LAXPC} observations of GRS 1915$+$105 in $3-60$ keV energy range in different variability classes. Results are obtained with combined data from {\it LAXPC10} and {\it LAXPC20}. CO and L$_i$ ($i=1,2,3,4$) denote the \texttt{constant} and multiple \texttt{Lorentzians} used to obtain the best fit. $\sigma$ denotes the significance of HFQPOs. $\rm HFQPO_{\rm rms}\%$ and $\rm Total_{\rm rms}\%$ represent the rms percentage of the HFQPO feature and the entire PDS. The centroid frequency (LC), FWHM (LW) and normalization (LN) of the detected HFQPOs are highlighted in bold font. All the errors are computed with 68\% confidence level. See text for details.}
	
	\resizebox{0.9\textwidth}{!}{%
		\begin{tabular}{l @{\hspace{0.4cm}} c @{\hspace{0.4cm}} c c @{\hspace{0.3cm}} c @{\hspace{0.3cm}} c @{\hspace{0.3cm}} c @{\hspace{0.3cm}} c @{\hspace{0.3cm}} c @{\hspace{0.4cm}} c @{\hspace{0.4cm}} c @{\hspace{0.4cm}} c @{\hspace{0.4cm}} c @{\hspace{0.4cm}} c @{\hspace{0.1cm}} c}
			
			\hline\hline
			& \multicolumn{6}{|c|}{Model Parameters} & & \multicolumn{3}{|c|}{Estimated Parameters} & \\
			
			\cline{2-7}
			\cline{9-11}\\
			
			MJD (Orbit) & CO  ($10^{-4}$) & & $\rm L_1$ & $\rm L_2$ & $\rm L_3$ & $\rm L_4$ & $\chi^{2}/dof$ & $\sigma$ & $\rm HFQPO_{\rm rms}\%$ & $\rm Total_{\rm rms}\%$ & Class & \\
			
			\hline
			& & LC & 0.0 & 0.0 & $5.05_{-0.07}^{+0.05}$ & $-$ & & & \\
			57451.89 (2351)$^{\dag}$ & $1.58_{-0.01}^{+0.01}$& LW & $5.47_{-0.43}^{+0.57}$ & $0.90_{-0.11}^{+0.14}$ & $1.09_{-0.12}^{+0.23}$ & $-$ & $170/229$ & $-$ & $-$ & $19.78\pm6.78$ & $\theta$ & \\
			& & LN & $0.0034_{-0.0007}^{+0.0007}$ & $ 0.033_{-0.004}^{+0.004}$& $ 0.0010_{-0.0001}^{+0.0001}$& $-$ & & & \\
			
			\hline
			& & LC & 0.0 & 0.0 & $3.25_{-0.01}^{+0.02}$ & $6.11_{-0.11}^{+0.08}$ & & & \\
			57452.82 (2365)$^{\dag}$& $1.34_{-0.01}^{+0.01}$ & LW & $3.94_{-0.49}^{+0.53}$ & $0.85_{-0.21}^{+0.24}$ & $0.44_{-0.01}^{+0.03}$ & $2.36_{-0.31}^{+0.38}$ & $168/232$ & $-$& $-$ & $17.71\pm3.59$ & $\chi$ & \\
			& & LN & $0.0037_{-0.0007}^{+0.0008}$ & $0.0055_{-0.0008}^{+0.0007}$ & $0.019_{-0.001}^{+0.002}$ & $0.00077_{-0.00004}^{+0.00005}$ & & & \\
			
			\hline
			& & LC & 0.0 & 0.0 & $6.19_{-0.16}^{+0.14}$ & $-$ & & & \\
			57453.35 (2373)$^{\dag}$& $1.53_{-0.01}^{+0.01}$& LW & $7.87_{-1.81}^{+3.23}$& $1.37_{-0.18}^{+0.19}$& $1.75_{-0.39}^{+0.61}$& $-$& $137/235$ & $-$& $-$& $20.09\pm4.43$& $\theta$& \\
			& & LN & $0.0011_{-0.0006}^{+0.0009}$ & $0.0303_{-0.0031}^{+0.0036}$ & $0.00046_{-0.00008}^{+0.00009}$ & $-$& & & \\
			
			\hline
			& & LC & 0.0 & $1.71_{-0.21}^{+0.20}$& $\textbf{71.52}_{-0.29}^{+0.36}$& $-$ & & & \\
			57504.02 (3124) & $1.86_{-0.01}^{+0.01}$& LW & $0.26_{-0.03}^{+0.04}$& $2.71_{-0.22}^{+0.23}$& $\textbf{1.25}_{-0.55}^{+0.62}$ & $-$& $151/233$& $3.67$ & $1.48 \pm 0.36$& $21.03 \pm 4.12$& $\omega$& \\
			& & LN & $0.21_{-0.04}^{+0.05}$& $0.00097_{-0.00013}^{+0.00016}$& $\textbf{0.00011}_{-0.00003}^{+0.00004}$& $-$& & & \\
			
			\hline
			& & LC & 0.0& 0.0& $0.75_{-0.05}^{+0.06}$& $2.73_{-0.18}^{+0.21}$ & & & \\
			57552.56 (3841)$^{\dag}$& $1.39_{-0.01}^{+0.01}$& LW & $8.45_{-2.25}^{+1.84}$& $0.33_{-0.05}^{+0.06}$& $0.77_{-0.11}^{+0.14}$& $4.05_{-1.24}^{+0.98}$& $165/231$& $-$& $-$& $7.64\pm2.18$ & $\delta$& \\
			& & LN & $0.00004_{-0.00001}^{+0.00001}$& $0.012_{-0.001}^{+0.002}$& $0.0014_{-0.0002}^{+0.0003}$& $0.00023_{-0.00004}^{+0.00004}$& & & \\
			\hline
			%\\
			& & LC & 0.0& $1.62_{-0.20}^{+0.23}$& $0.83_{-0.05}^{+0.04}$& \textbf{$\textbf{69.31}_{-0.16}^{+0.11}$} & & & \\
			57553.88 (3860)& $1.36_{-0.01}^{+0.01}$& LW & $0.018_{-0.004}^{+0.005}$& $3.10_{-0.16}^{+0.19}$& $1.05_{-0.19}^{+0.24}$& \textbf{$\textbf{1.39}_{-1.06}^{+0.94}$}& $134/228$& $2.75$& $1.54\pm0.48$& $9.69\pm2.23$ & $\delta$& \\
			& & LN & $0.29_{-0.02}^{+0.03}$& $0.0012_{-0.0003}^{+0.0003}$& $0.0021_{-0.0003}^{+0.0003}$& \textbf{$\textbf{0.00011}_{-0.00004}^{+0.00004}$}& & & \\
			\hline
			
			%\\
			& &  {LC} &  {0.0}& $ {0.0}$& $ {3.59_{-0.34}^{+0.32}}$& $ {3.35_{-0.28}^{+0.39}}$ & & & \\
			57689.10 (5862)$^{\dag}$ & $ {1.72_{-0.01}^{+0.01}}$&  {LW} & $ {0.023_{-0.002}^{+0.001}}$& $ {1.34_{-0.14}^{+0.15}}$& $ {6.41_{-0.46}^{+0.49}}$ & $ {0.98_{-0.23}^{+0.26}}$ & $ {129/231}$& $-$& $-$& $ {26.30\pm6.12}$& $ {\beta}$& \\
			& &  {LN} & $ {5.05_{-1.21}^{+1.45}}$& $ {0.013_{-0.002}^{+0.002}}$& $ {0.00071_{-0.00008}^{+0.00008}}$& $ {0.0012_{-0.0003}^{+0.0003}}$ & & & \\
			\hline
			
			& & LC & 0.0& 0.0& $1.73_{-0.26}^{+0.33}$& $-$& & & \\
			57705.22 (6102)$^{\dag}$& $1.37_{-0.01}^{+0.01}$& LW & $1.68_{-0.61}^{+0.71}$& $0.06_{-0.01}^{+0.02}$& $4.21_{-0.16}^{+0.15}$ & $-$& {  $154/231$}& $-$& $-$&  {$11.68\pm2.71$}& $\delta$& \\
			& & LN & $0.0019_{-0.0003}^{+0.0006}$& $0.16_{-0.03}^{+0.08}$& $0.00092_{-0.00023}^{+0.00021}$& $-$& & & \\
			\hline
			%\\
			& & LC & 0.0& 0.0& $5.14_{-0.02}^{+0.04}$& $10.33_{-0.21}^{+0.18}$& & & \\
			57891.88 (8863)$^{\dag}$& $1.48_{-0.01}^{+0.01}$& LW & $13.47_{-1.66}^{+1.41}$& $0.18_{-0.03}^{+0.04}$& $1.15_{-0.11}^{+0.12}$& $3.47_{-0.89}^{+1.43}$& {  $183/232$}& $-$& $-$&   {$27.38\pm5.69$} & $\rho \textprime$ & \\
			& & LN & $0.0011_{-0.0001}^{+0.0001}$& $0.42_{-0.12}^{+0.16}$& $0.0032_{-0.0002}^{+0.0003}$& $0.00042_{-0.00004}^{+0.00006}$& & & \\
			\hline
			%\\
			& & LC & 0.0& 0.0& $5.74_{-0.05}^{+0.06}$& $10.63_{-0.23}^{+0.28}$& & & \\
			57892.74 (8876)$^{\dag}$& $1.58_{-0.01}^{+0.01}$& LW & $3.80_{-0.67}^{+0.81}$& $0.13_{-0.03}^{+0.05}$& $1.54_{-0.14}^{+0.15}$& $8.43_{-0.93}^{+1.18}$& {  $144/222$}& $-$& $-$&   {$23.53\pm2.70$} & $\rho$& \\
			& & LN & $0.0016_{-0.0003}^{+0.0004}$& $0.71_{-0.18}^{+0.21}$& $0.0019_{-0.0001}^{+0.0002}$& $0.00043_{-0.00002}^{+0.00005}$& & & \\
			\hline
			%\\
			& & LC & 0.0& $-$& $-$& $-$& & & \\
			57943.69 (9629)$^{\dag}$& $1.52_{-0.01}^{+0.01}$& LW & $0.14_{-0.01}^{+0.02}$& $-$& $-$& $-$& {  $149/239$} & $-$& $-$&  { $35.49\pm5.35$ }& $\kappa$& \\
			& & LN & $1.26_{-0.24}^{+0.28}$& $-$& $-$& $-$& & & \\
			\hline
			%\\
			& & LC & 0.0& 0.0& \textbf{$\textbf{ {71.55}}_{-0.96}^{+0.96}$}& $-$& & & \\
			57943.69 (9633)& $1.59_{-0.01}^{+0.01}$& LW & $ {0.13_{-0.01}^{+0.01}}$ & $ {2.21_{-0.13}^{+0.15}}$& \textbf{$\textbf{ {5.67}}_{-2.17}^{+2.76}$}& $-$&$ {172/234}$& $ {3.71}$& $ {2.14\pm0.50}$&   {$38.32\pm6.67$} & $\kappa$& \\
			& & LN & $ {1.36_{-0.21}^{+0.24}}$& $ {0.016_{-0.002}^{+0.002}}$& \textbf{$\textbf{0.000052}_{-0.000014}^{+0.000019}$}& $-$& & \\
			\hline
			%\\
			& & LC & 0.0& \textbf{$\textbf{69.76}_{-0.74}^{+0.89}$}& $-$& $-$& & & \\
			57946.10 (9666)& $1.55_{-0.01}^{+0.01}$& LW & $0.17_{-0.01}^{+0.02}$& \textbf{$\textbf{5.01}_{-2.29}^{+2.93}$}& $-$& $-$& {  $146/237$}& $ {3.27}$& $2.55\pm0.71$&   {$37.28\pm6.31$} & $\kappa$& \\
			& & LN & $1.12_{-0.22}^{+0.26}$& \textbf{$\textbf{0.000085}_{-0.000026}^{+0.000042}$}& $-$& $-$& & & \\
			\hline
			%\\	
			& & LC & 0.0& $6.79_{-1.05}^{+1.38}$& \textbf{$\textbf{70.45}_{-0.53}^{+0.54}$}& $-$& & & \\
			57946.34 (9670)& $1.67_{-0.01}^{+0.01}$& LW & $0.13_{-0.01}^{+0.01}$& $7.43_{-1.98}^{+1.69}$& \textbf{$\textbf{4.67}_{-1.07}^{+1.56}$}& $-$& {  $142/234$}& $ {5.06}$& $2.49\pm0.37$&  {$37.98\pm5.35$} & $\kappa$& \\
			& & LN & $1.55_{-0.22}^{+0.25}$& $0.000078_{-0.000014}^{+0.000021}$& \textbf{$\textbf{0.000086}_{-0.000017}^{+0.000018}$}& $-$& & & \\
			\hline
			%\\
			& & LC & 0.0& $-$& $-$& $-$& & & \\
			57961.39 (9891)$^{\dag}$& $5.46_{-0.01}^{+0.01}$& LW & $0.29_{-0.02}^{+0.04}$& $-$& $-$& $-$& {  $147/240$}& $-$& $-$&  {$31.51\pm4.35$} & $\kappa$& \\
			& & LN & $0.44_{-0.08}^{+0.12}$& $-$& $-$& $-$& & & \\
			\hline
			%\\
			& & LC & 0.0& 0.0& $2.14_{-0.42}^{+0.48}$& \textbf{$\textbf{71.68}_{-0.75}^{+0.84}$}& & & & \\
			57961.39 (9894)& $1.57_{-0.01}^{+0.01}$& LW & $2.14_{-0.01}^{+0.02}$& $0.09_{-0.02}^{+0.03}$& $1.25_{-0.31}^{+0.36}$& \textbf{$\textbf{4.70}_{-2.21}^{+2.60}$}& {  $156/232$}& $ {3.43}$& $2.28\pm0.63$&  {$42.31\pm8.66$} & $\kappa$& \\
			& & LN & $0.04_{-0.01}^{+0.01}$& $1.70_{-0.53}^{+0.84}$& $0.0042_{-0.0011}^{+0.0011}$& \textbf{$\textbf{0.000072}_{-0.000021}^{+0.000038}$}& & & & & \\
			\hline 
			%\\
			& & LC & 0.0& $ {0.72_{-0.01}^{+0.02}}$& \textbf{$\textbf{70.47}_{-0.65}^{+0.69}$}& $-$ & & & \\
			57961.59 (9895)& $2.11_{-0.01}^{+0.01}$& LW & $ {0.13_{-0.01}^{+0.02}}$ & $ {4.01_{-0.62}^{+0.66}}$& \textbf{$\textbf{3.86}_{-1.32}^{+1.69}$}& $-$& $ {171/234}$& $ {4.40}$& $1.98\pm0.41$& $ {39.38\pm7.30}$& $\kappa$& \\
			& & LN & $ {1.59_{-0.28}^{+0.36}}$& $ {0.00105_{-0.00033}^{+0.00069}}$& \textbf{$\textbf{0.000066}_{-0.000015}^{+0.000017}$}& $-$ & & & \\
			\hline
			%\\
			& & LC & 0.0 & $0.32_{-0.02}^{+0.04}$& $6.79_{-0.24}^{+0.34}$& \textbf{$\textbf{68.14}_{-0.61}^{+0.65}$}& & & \\
			57995.30 (10394)& $2.06_{-0.01}^{+0.01}$& LW & $ {0.08_{-0.01}^{+0.02}}$& $0.61_{-0.02}^{+0.03}$& $2.46_{-1.10}^{+0.91}$& \textbf{$\textbf{5.42}_{-1.11}^{+1.49}$}& $ {165/231}$& $ {5.55}$& $2.25\pm0.31$&  {$34.27\pm9.76$}& $\omega$& \\
			& & LN & $ {1.23_{-0.39}^{+0.59}}$ & $0.05_{-0.01}^{+0.01}$& $0.00012_{-0.00003}^{+0.00004}$& \textbf{$\textbf{0.000061}_{-0.000011}^{+0.000012}$}& & & \\
			\hline	
			%\\
			& & LC & 0.0& $0.11_{-0.03}^{+0.02}$& \textbf{$\textbf{70.01}_{-0.49}^{+0.56}$}& $-$& & & \\
			57996.46 (10411)& $1.75_{-0.01}^{+0.01}$& LW & $0.03_{-0.01}^{+0.01}$& $0.57_{-0.05}^{+0.07}$& \textbf{$\textbf{5.96}_{-0.99}^{+1.22}$}& $-$& {  $193/234$}& $ {7.01}$& $2.66\pm0.29$&  {$26.45\pm7.47$}& $\omega$& \\
			& & LN & $1.75_{-0.23}^{+0.44}$& $0.08_{-0.01}^{+0.01}$& \textbf{$\textbf{0.000077}_{-0.000011}^{+0.000012}$}& $-$& & & \\
			\hline
			%\\
			& & LC & 0.0& $ {1.89_{-0.31}^{+0.23}}$& \textbf{$\textbf{72.22}_{-0.39}^{+0.49}$}& $-$ & & & \\
			58007.80 (10579)& $1.83_{-0.01}^{+0.01}$& LW & $ {0.25_{-0.03}^{+0.05}}$& $ {3.04_{-0.40}^{+0.42}}$& \textbf{$\textbf{2.25}_{-1.48}^{+0.77}$}& $-$& $ {128/234}$& $ {4.95}$& $1.86\pm0.37$&  {$19.01\pm3.66$} & $\omega$& \\
			& & LN & $ {0.18_{-0.03}^{+0.07}}$& $ {0.00068_{-0.00009}^{+0.00011}}$& \textbf{$\textbf{0.000099}_{-0.000020}^{+0.000032}$}& $-$& & & \\
			\hline
			%\\
			& & LC & 0.0& 0.0& $2.25_{-0.05}^{+0.06}$& \textbf{$\textbf{72.32}_{-0.21}^{+0.23}$}& & & \\
			58008.08 (10583)& $1.37_{-0.01}^{+0.01}$& LW & $9.71_{-2.66}^{+4.19}$& $0.27_{-0.03}^{+0.03}$& $1.52_{-0.21}^{+0.23}$& \textbf{$\textbf{3.61}_{-0.45}^{+0.48}$}& {  $133/231$}& $ {11.00}$& $2.46\pm0.19$&  {$12.98\pm3.72$ }& $\gamma$& \\
			& & LN & $0.00021_{-0.00009}^{+0.00016}$& $0.062_{-0.011}^{+0.013}$& $0.0010_{-0.0001}^{+0.0001}$& \textbf{$\textbf{0.00011}_{-0.00001}^{+0.00001}$}& & & \\
			\hline
			%\\
			& & LC & 0.0& 0.0& $0.61_{-0.02}^{+0.04}$& $-$& & & \\
			58046.36 (11154)$^{\dag}$& $1.36_{-0.01}^{+0.01}$& LW & $4.66_{-0.32}^{+0.32}$& $0.21_{-0.04}^{+0.05}$& $0.43_{-0.06}^{+0.09}$& $-$& {  $166/234$}& $-$& $-$&  {$7.91\pm1.77$}& $\delta$& \\
			& & LN & $0.00086_{-0.00009}^{+0.00011}$& $0.013_{-0.003}^{+0.004}$& $0.0018_{-0.0002}^{+0.0003}$& $-$& & & \\
			
			\hline
			& & LC &  {0.0}&  {0.0}& $ {3.26_{-0.02}^{+0.02}}$& $ {6.32_{-0.06}^{+0.07}}$& & & \\
			 {58209.13 (13559)}$^{\dag}$& $ {1.43_{-0.01}^{+0.01}}$&  {LW} & $ {6.18_{-0.35}^{+0.42}}$& $ {0.84_{-0.11}^{+0.13}}$& $ {0.64_{-0.04}^{+0.04}}$& $ {1.24_{-0.21}^{+0.23}}$& $ {171/232}$& $-$& $-$& $ {24.02\pm4.23}$& $ {\chi}$& \\
			& &  {LN} & $ {0.0059_{-0.0008}^{+0.0006}}$& $ {0.018_{-0.002}^{+0.002}}$& $ {0.016_{-0.001}^{+0.001}}$& $ {0.00104_{-0.00011}^{+0.00012}}$& & & \\
			
			\hline
			& &  {LC} &  {0.0}&  {0.0}& $ {2.20_{-0.01}^{+0.01}}$& $ {4.42_{-0.03}^{+0.03}}$& & & \\
			 {58565.82 (18839)}$^{\dag}$& $ {1.52_{-0.01}^{+0.01}}$&  {LW} & $ {7.76_{-0.81}^{+1.29}}$& $ {0.57_{-0.15}^{+0.27}}$& $ {0.15_{-0.01}^{+0.03}}$& $ {0.39_{-0.09}^{+0.11}}$& $ {130/232}$& $-$& $-$& $  {21.14\pm5.13}$& $ {\chi}$& \\
			& &  {LN} & $ {0.0045_{-0.0009}^{+0.0007}}$& $ {0.014_{-0.002}^{+0.003}}$& $ {0.031_{-0.003}^{+0.004}}$& $ {0.0048_{-0.0008}^{+0.0009}}$& & & \\
			\hline
			
		\end{tabular}%
	}
	\label{table:PDS_parameters}
	
	\begin{list}{}{}
		\item $^{\dagger}$ Non-detection of HFQPO.
	\end{list}
\end{table*}

\section{Timing Analysis and Results}

\label{s:timing}

\subsection{Variability Classes and Color-Color Diagram (CCD)}

We generate $1$ s binned light curve in the energy range $3-60$ keV, after combining the data from {\it LAXPC10} and {\it LAXPC20} while studying the structured variability in different classes. Following \cite{Agrawal-etal2018, Sreehari-etal2019, Sreehari-etal2020}, we correct the dead-time effect in all the light curves and calculate the average incident and detected count rates in $3-60$ keV energy range as tabulated in Table \ref{table:Obs_details}. The light curves are generated in $3-6$ keV, $6-15$ keV and $15-60$ keV energy ranges to plot the CCDs. We define the soft color and hard color following \cite{Sreehari-etal2020} as ${\rm HR1}=B/A$ and ${\rm HR2}=C/A$, where $A$, $B$ and $C$ are the photon count rates in $3 - 6$ keV, $6 - 15$ keV and $15 - 60$ keV energy bands, respectively. The CCDs are obtained by plotting HR1 against HR2. It may be noted that the combined {\it LAXPC10} and {\it LAXPC20} background count rate ($\sim 225$ cts/s) is negligible ($\sim 2.5\%$) compared to the combined source count rate ($\sim 10000$ cts/s). Accordingly, we incorporate the background correction while carrying out the color-color and spectral analyses. However, power spectra are generated without background correction of light curves.

\begin{figure}
	\begin{center}
		\includegraphics[width=\columnwidth]{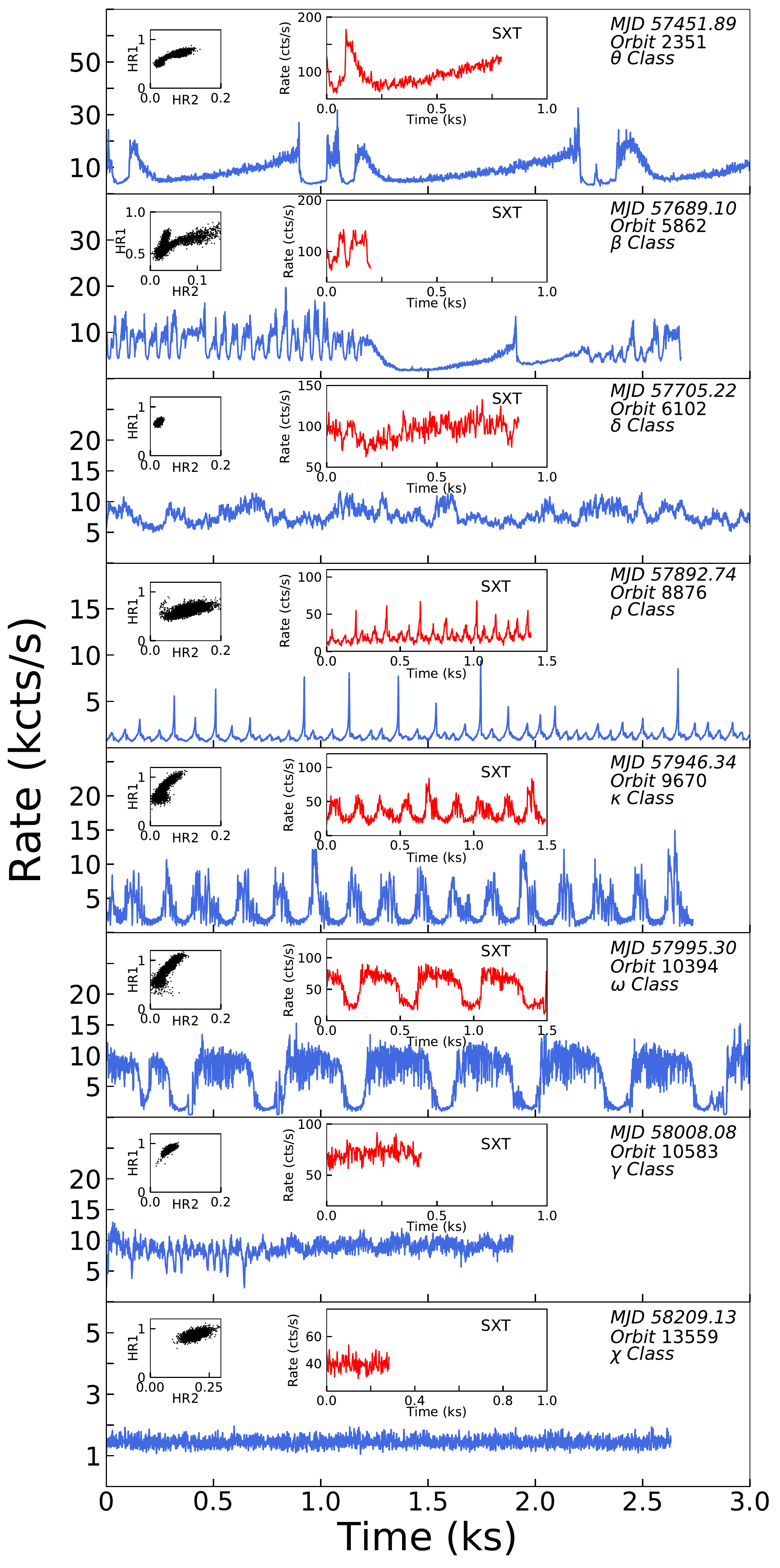}
	\end{center}

	\caption{Background subtracted and dead-time corrected $1$ s binned light curves of GRS 1915+105 observed with {\it AstroSat}. Light curves corresponding to eight different variability classes, namely  $\theta$, $\beta$, $\delta$, $\rho$, $\kappa$, $\omega$, $\gamma$ and  $\chi$ are depicted from top to bottom panels. Each light curve is obtained by combining {\it LAXPC10} and {\it LAXPC20} data in $3-60$ keV energy band. In every panel, the CCD and the {\it SXT} light curves ($0.5-7$ keV) are also shown at the top-left and top-middle insets, respectively. See text for details.
	}
	\label{fig:lcurvCCD}
\end{figure}

Following the classification scheme of \cite{Belloni-etal2000}, the light curves and CCDs indicate the presence of seven `softer' variability classes as $\theta$, $\beta$, $\delta$, $\rho$, $\kappa$, $\omega$ and $\gamma$ with an additional variant ($\rho \textprime$) of $\rho$ class \citep{Athulya-etal2021}, and one `harder' variability class ($\chi$). The hardness ratios $HR1$ and $HR2$ for each class are tabulated in Table \ref{table:Obs_details}. It is clearly seen that the soft color $HR1$ and the hard color $HR2$ are generally less than $0.9$ and $0.1$, respectively, which imply the `softer' variability classes. In Fig. \ref{fig:lcurvCCD}, we present background subtracted and dead-time corrected {\it LAXPC} light curves of eight different variability classes ($\theta$, $\beta$, $\delta$, $\rho$, $\kappa$, $\omega$, $\gamma$ and $\chi$) with the CCD at the top-left inset of each panel. Different structured variability patterns along with the variations in the CCDs are observed in various classes. 

During the {\it AstroSat} campaign, the source was initially observed in $\theta$ variability class \citep[see][]{Banerjee-etal2021}, where the count rate went up to $20$ kcts/s and both colors are softened as observed in the CCD. Further, the source was found in $\omega$, $\delta$ and $\beta$ variability classes (see Table \ref{table:Obs_details}.) Subsequently, the source was found in a variability class $\rho$ on MJD 57892.74. In this period, we see a `flare' like nature in the light curve, and in the CCD, the points are distributed more towards the harder range. Next, the source displayed $\kappa$ class variability, in which the count rates were high as 15 kcts/s and multiple `dips' (low counts) of few tens to hundred seconds of duration are observed. In addition, small duration ($\sim$ few seconds) `non-dips' (high counts) are also present between the two `dips'. The CCD shows a uniform C-shaped distribution of points. In $\omega$ class, the duration of `non-dips' between two `dips' increases up to a few hundred of seconds, and the CCD shows a similar pattern as observed in $\kappa$ class. In $\gamma$ class, the long duration `dips' are absent instead of small `dips' with a few seconds of duration are observed along with high counts. A diagonally elongated distribution of points is found in the CCD. Finally, the source was found in `harder' variability class ($\chi$) with count rate less than $1.5$ kcts/s \citep{Athulya-etal2021}. In each panel of Fig. \ref{fig:lcurvCCD}, we show the {\it SXT} light curves of the same observations in the energy range of $0.5-7$ keV at the top-middle inset. Similar structured variabilities are also observed in {\it SXT} light curves as seen in {\it LAXPC} observations.

%%%%%%%%%%%%%%%%%%%%%%Figure%%%%%%%%%%%%%%%%%%%%%%%%%%%%%%%%%%%%%%%%%%%%%%%

%%%%%%%%%%%%%%%%%%% Figure %%%%%%%%%%%%%%%%%%%%%%%%%%%
\begin{figure}
	\begin{center}
		\includegraphics[width=\columnwidth]{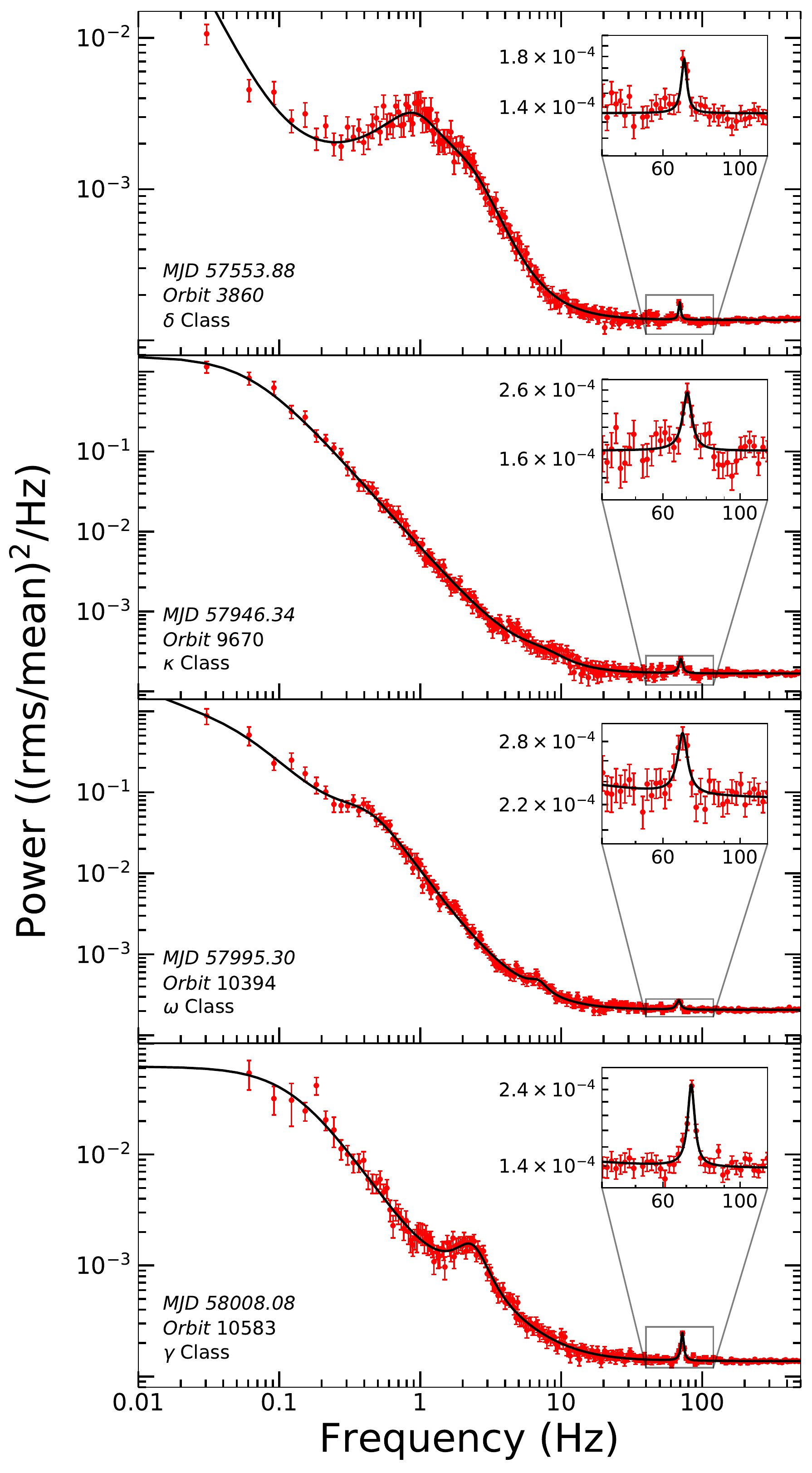}
	\end{center}
	
	\caption{Power density spectra of four different variability classes in the presence of HFQPOs. Each PDS is obtained in $3-60$ keV energy range by combining {\it LAXPC10} and {\it LAXPC20} observations. In each panel, variability classes are marked, and the detected HFQPO features (zoomed view) are shown in the inset. See text for details.}
	\label{fig:PDS_QPO}
\end{figure}
%%%%%%%%%%%%%%%%%%% Figure %%%%%%%%%%%%%%%%%%%%%%%%%%%%%

\subsection{Static Power Spectra}
\label{s:sps}

We generate light curves of $1$ ms resolution corresponding to each variability class with combined data from {\it LAXPC10} and {\it LAXPC20}. We generate a power density spectrum (PDS) for each observation considering Nyquist frequency of $500$ Hz with these light curves. We choose $32768$ bins per interval for generating the respective PDS, which are further averaged to obtain the final PDS. A geometric binning factor of $1.03$ is used for the power spectral analysis.  Finally, the dead-time corrected power spectra are obtained following \cite{Agrawal-etal2018,Sreehari-etal2019,Sreehari-etal2020}.

Each PDS (in units of $\rm (rms/mean)^{2}/Hz$) is then modelled using multiple \texttt{Lorentzians} and a \texttt{constant} component in the wide frequency range of $0.01-500$ Hz. Each \texttt{Lorentzian} is represented by three parameters, namely centroid (LC), width (LW), and normalization (LN). In Fig. \ref{fig:PDS_QPO}, we present the model fitted PDS of $\delta$, $\kappa$, $\omega$ and $\gamma$ variability classes with the detected HFQPO feature as depicted in the inset of each panel. The variability class and the observation details are marked in each panel. First, we begin with the modeling of the power spectrum of $\gamma$ class observation using the model combination of a \texttt{constant} and four \texttt{Lorentzians}. Initially, two zero centroid \texttt{Lorentzians} and one \texttt{Lorentzian} with centroid frequency at 2.25 Hz along with a \texttt{constant} component are used to fit the entire PDS. The fit is resulted in a $\chi_{\rm red}^{2}$ of $310/234=1.32$.  Further, to model the HFQPO feature, we include an additional \texttt{Lorentzian} with the centroid frequency of $\sim 72$ Hz. We emphasize that while modeling the entire PDS and estimating the errors associated with the model parameters, all the model parameters are kept free. The best fit is obtained with a $\chi_{\rm red}^{2}$ of $133/231=0.58$ \cite[see also][]{Sreehari-etal2020}. We follow the above procedure to fit the PDS in the presence of HFQPO features for $\delta$, $\kappa$ and $\omega$ classes, and the best fitted model parameters along with errors are tabulated in Table \ref{table:PDS_parameters}. In addition, we also note the presence of a broad feature in the PDS of $\delta$ and $\gamma$ class observations at $\sim 1$ Hz and $2$ Hz, respectively, which is absent in $\kappa$ and $\omega$ classes.

%%%%%%%%%%%%%%%%%%%%%%%%%%%%%% Figure-4 %%%%%%%%%%%%%%%%%%%%%%%%%%%%%%
\begin{figure}

	\begin{center}
	\includegraphics[width=\columnwidth]{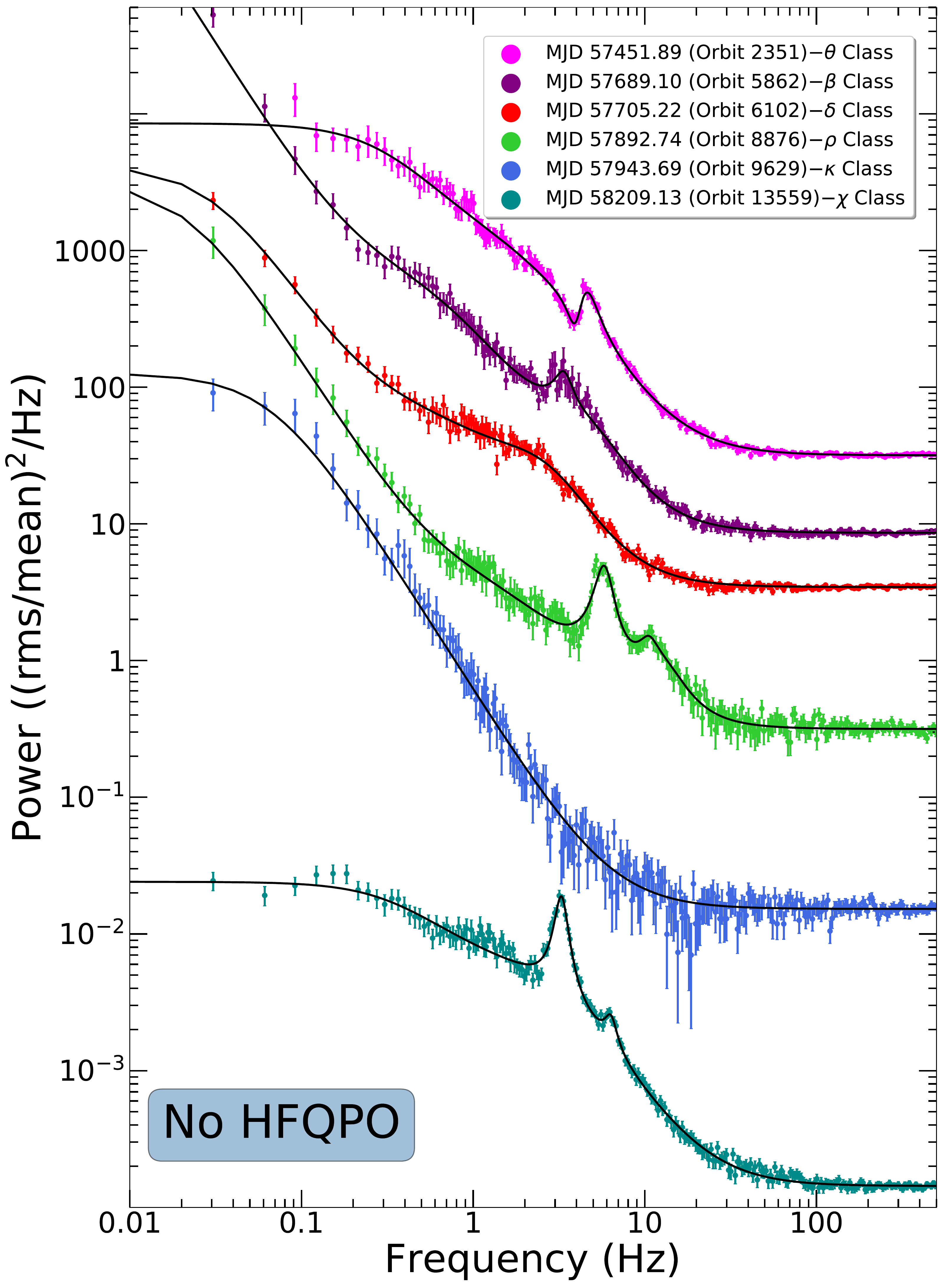}
	\end{center}

	\caption{Power density spectra of six different observations in $3-60$ keV energy band, where HFQPO is absent. Magenta, purple, red, green, blue and darkcyan filled circles denote the PDS corresponding to  $\theta$, $\beta$, $\delta$, $\rho$, $\kappa$ and $\chi$ classes, respectively. The power corresponding to $\theta$, $\beta$, $\delta$, $\rho$ and $\kappa$ class observations are scaled by multiplying $200000$, $50000$, $25000$, $2000$ and $100$ for the purpose of clarity. See text for details.
	}
	\label{fig:PDS_noQPO}
\end{figure}
%%%%%%%%%%%%%%%%%%%%%%%%%%%%%% Figure-4 %%%%%%%%%%%%%%%%%%%%%%%%%%%%%%

The presence of a HFQPO feature in the power spectra is determined by means of quality factor ($Q = LC/LW \ge 3$) and significance ($\sigma = LN/err_{\rm neg} \ge 3$), where $LC$, $LW$, $LN$ and $err_{\rm neg}$ denote the centroid frequency, width, normalization, and negative error of normalization of the fitted \texttt{Lorentzian} function \cite[and references therein]{Sreehari-etal2020}. Best fitted PDS of $\gamma$ class observation shows a strong signature of HFQPO feature of centroid frequency $72.32_{-0.21}^{+0.23}$ Hz with significance of $11\sigma$ as shown in the inset of the bottom panel of Fig. \ref{fig:PDS_QPO}. We calculate the percentage rms of the HFQPO by taking the square root of the definite integral of \texttt{Lorentzian} function \citep[and references therein]{VanderKlis1989,Ribeiro-etal2019,Sreehari-etal2020} with the HFQPO fitted parameters (\textit{i.e.,} LC, LW and LN) and obtain as $2.46\pm 0.19\%$. Further, we calculate the percentage rms of the entire PDS in the wide frequency range $0.01-500$ Hz and for $\gamma$ class observation, we obtain its value as  $12.98\pm3.72\%$. We detect $11$ such HFQPO features in four variability classes ($i.e.$, $\delta$, $\kappa$, $\omega$, $\gamma$) with centroid frequencies in the range of $68.14-72.32$ Hz, significance in the range of $2.75-11.0\sigma$ and the percentage rms in the range of $1.48-2.66$. The model fitted as well as estimated parameters of all observations are tabulated in Table \ref{table:PDS_parameters}.

To address the non-detection of the HFQPO features, we compute the percentage rms amplitude ($rms\%$) of the most significant peak near the observed HFQPO frequency for all the observations. We find the highest value of the percentage rms amplitude as $0.94 \pm 0.15$ ($rms\%$ less than this value considered as `insignificant QPO rms') for non-detection of HFQPO feature (see Table \ref{table:noHFQPO}, Fig. \ref{fig:PDS_noQPO} for PDS of $\kappa$ class). On contrary, we obtain the lowest value of $rms\%$ for the confirmed HFQPO detection as $1.48 \pm 0.36$ in the power spectra. Accordingly, we identify thirteen observations in $\theta$, $\beta$, $\delta$, $\rho$, $\kappa$ and $\chi$ variability classes that do not exhibit the signature of HFQPO feature as presented in Table \ref{table:noHFQPO}. Further, for reconfirmation \citep{Belloni-etal2001, Sreehari-etal2020}, we include one additional \texttt{Lorentzian} in the PDS of Orbit 9629 ($\kappa$ class, as an example) by freezing the centroid frequency at $69.76$ Hz and width at $5.01$ Hz, similar to the HFQPO characteristics obtained in Orbit 9666 ($\kappa$ class) having similar exposure. The significance of the best fitted \texttt{Lorentzian} feature is found to be $1.21$ at $1 \sigma$ unit, which indicates the non-detection of HFQPO. All the model fitted parameters of the power spectra are presented in Table \ref{table:PDS_parameters}. In Fig. \ref{fig:PDS_noQPO}, we show the model fitted PDS corresponding to $\theta$, $\beta$, $\delta$, $\rho$, $\kappa$ and $\chi$ classes, modelled with \texttt{Lorentzians} and a \texttt{constant}. Note that in wide frequency band, the power spectra of $\delta$ class observation (Orbit 6102) shows a bump like feature near $2$ Hz, whereas the PDS corresponding to $\kappa$ class (Orbit 9629) is completely featureless. We find a strong Type-C LFQPO at $\sim 5.05$ Hz, $\sim 5.74$ Hz and $\sim 3.2$ Hz for $\theta$, $\rho$ and $\chi$ classes, respectively, whereas LFQPO feature at $\sim 3.35$ Hz is found broader for $\beta$ variability class. Here, we avoid the discussion on LFQPOs as it is beyond the scope of the present work.

%%%%%%%%%%%%%%%%%%%%%%%%%%%% Figure-5 %%%%%%%%%%%%%%%%%%%%%%%%%%%%
\begin{figure}
	\begin{center}
		\includegraphics[width=\columnwidth]{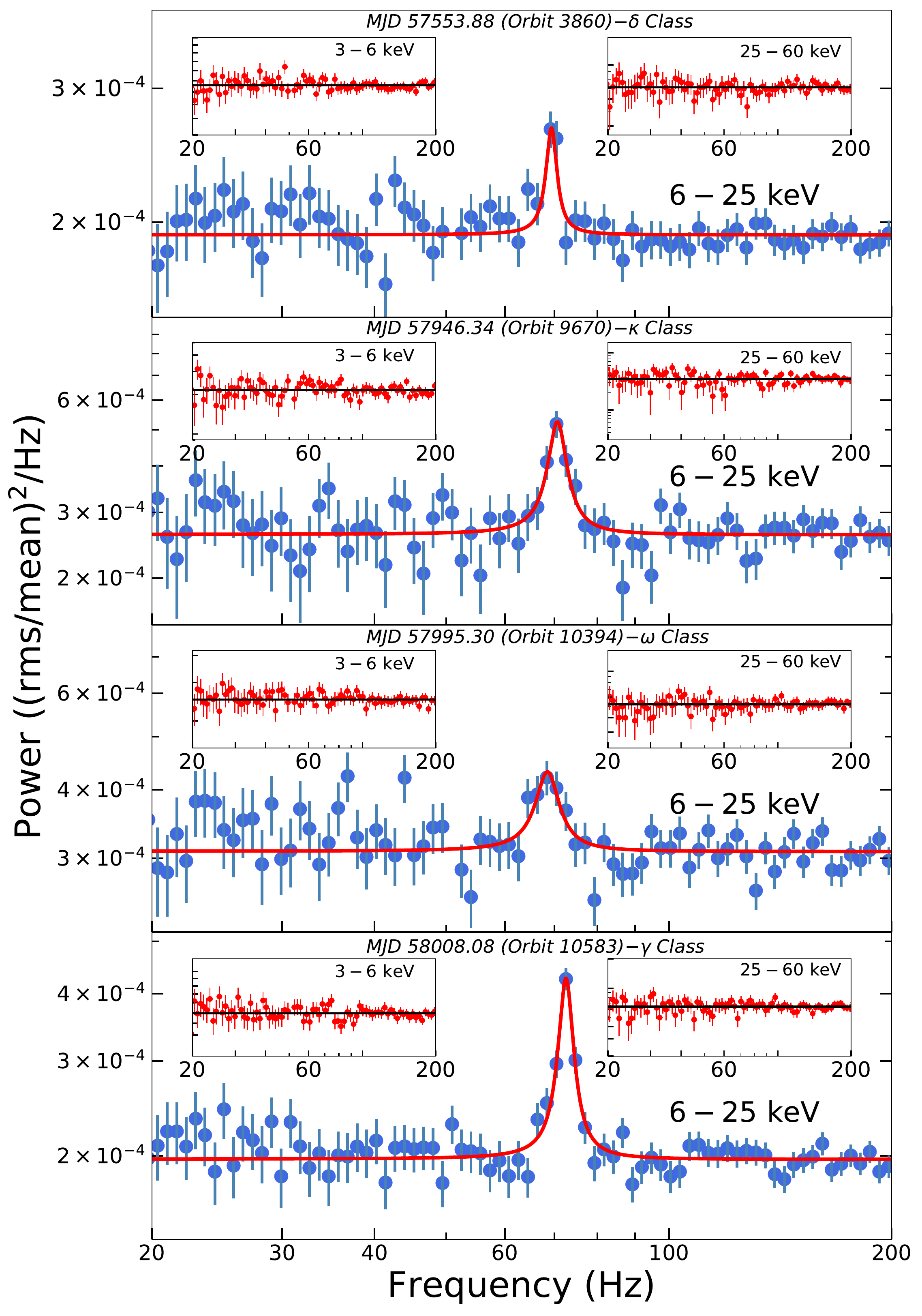}
	\end{center}
	
	\caption{Energy dependent power density spectra depicted in $20-200$ Hz frequency range. PDS corresponding to $\delta$, $\kappa$, $\omega$ and $\gamma$ class observations are presented in sequence from top to bottom panels. Energy ranges are marked in each panel including insets. The PDS corresponding to $3-6$ keV and $25-60$ keV energy ranges are modelled with a \texttt{constant}, whereas the PDS in $6-25$ keV energy range is modelled using a \texttt{constant} and a \texttt{Lorentizian}. See text for details.}
	\label{fig:PDS_zoom}
\end{figure} 
%%%%%%%%%%%%%%%%%%%%%%%%%%%% Figure-5 %%%%%%%%%%%%%%%%%%%%%%%%%%%%
 
%%%%%%%%%%%%%%%%%%%%%%%%%%%% Figure-6 %%%%%%%%%%%%%%%%%%%%%%%%%%%%
 \begin{figure*}
 	\begin{center}
 		\includegraphics[height=6cm, width=\columnwidth]{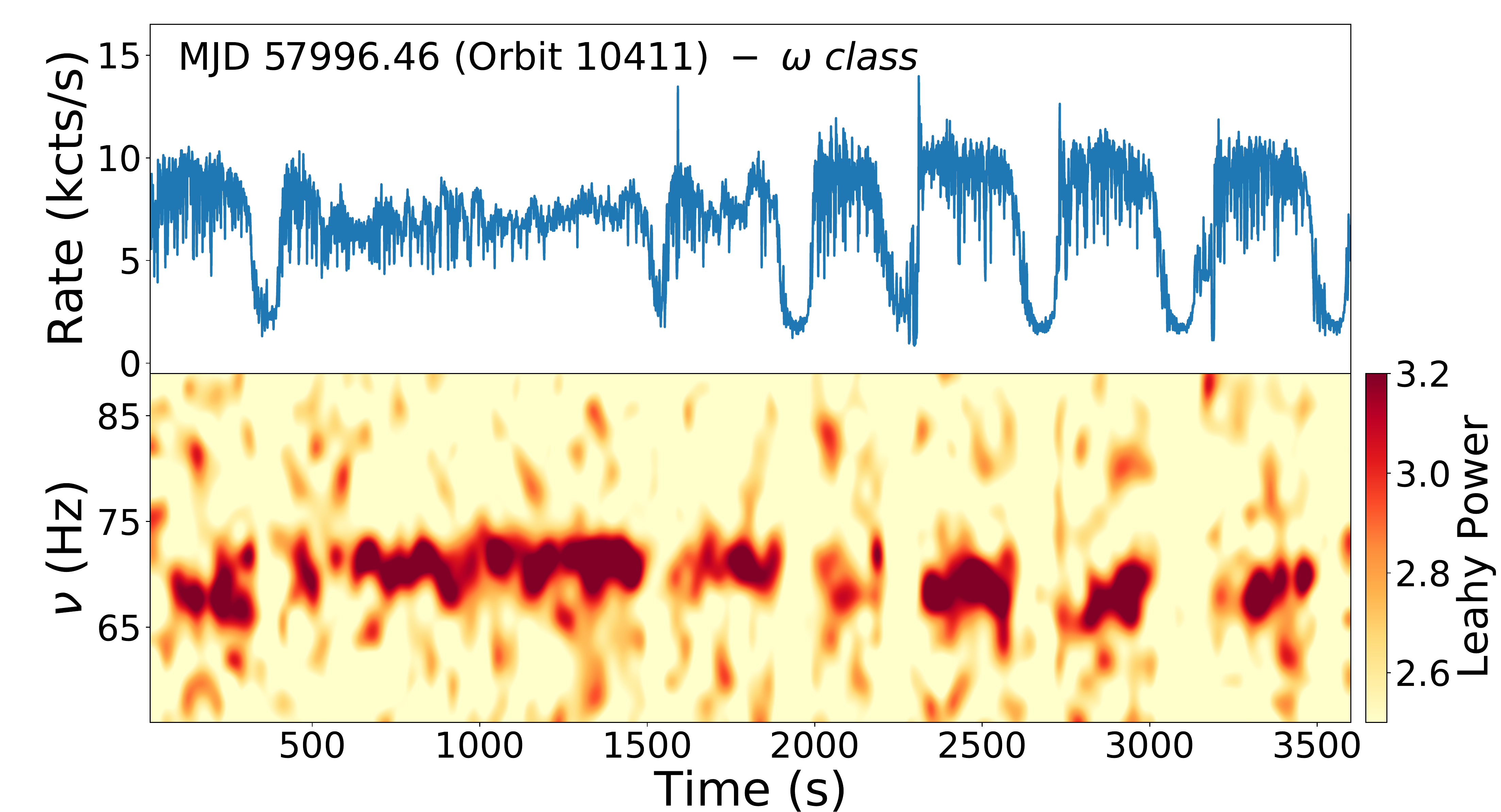}
 		\includegraphics[height=6cm, width=\columnwidth]{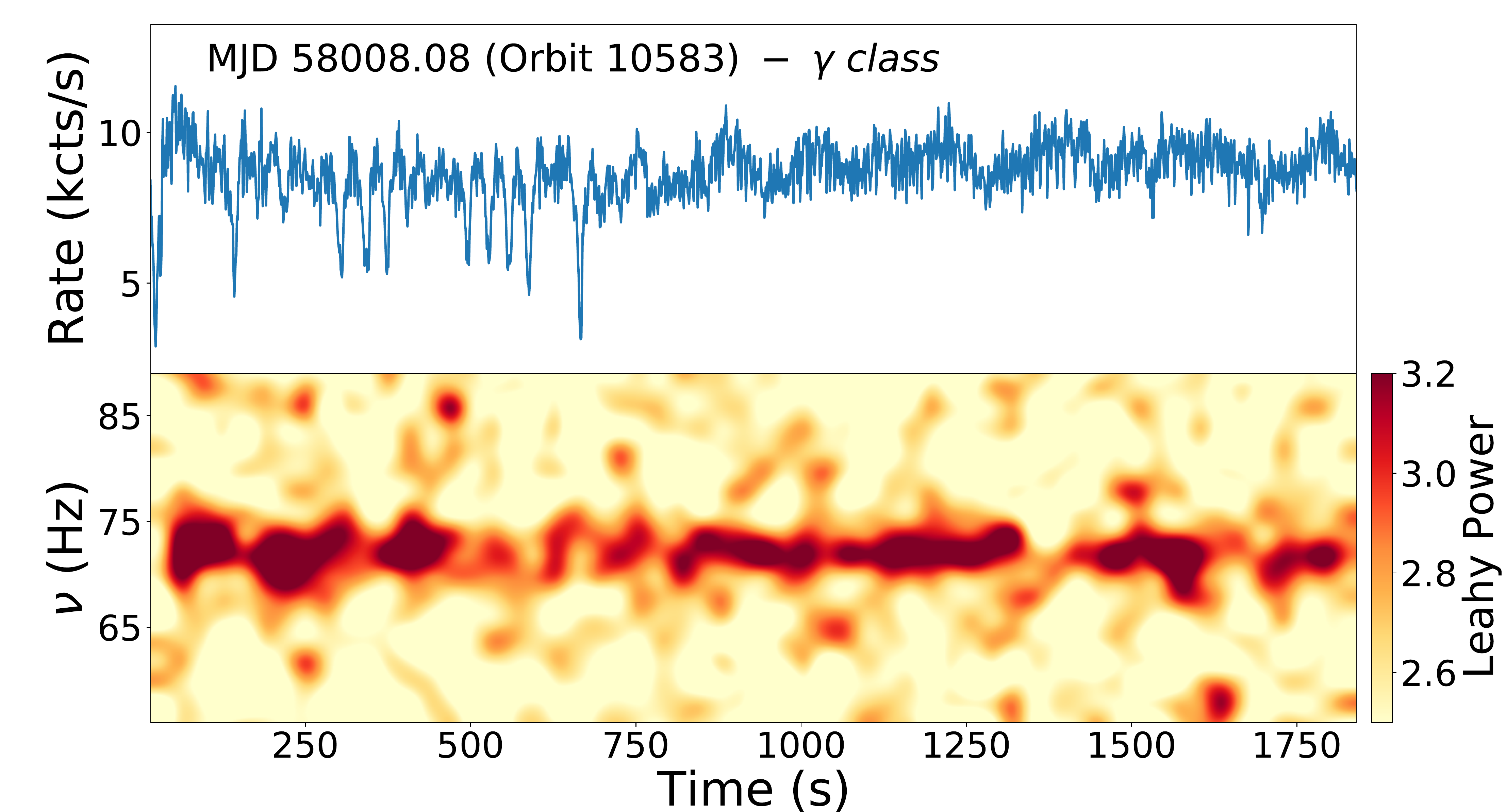}
 	\end{center}	
 	\caption{The light curves ($3 - 60$ keV) of $\omega$ and $\gamma$ classes are depicted in the top panels of each plot. In the bottom panels of each plot, the dynamical power spectra generated from the high resolution ($1$ ms) light curve are presented. Here, $32$ s segment size and $2$ Hz frequency bin are used to represent the dynamical power spectra. The obtained results are presented using color code, where colorbars are marked in the right of the bottom panels. See text for details.}
 	\label{fig:Dyn_PDS}
 \end{figure*}
%%%%%%%%%%%%%%%%%%%%%%%%%%%% Figure-6 %%%%%%%%%%%%%%%%%%%%%%%%%%%%
 
\subsection{Energy dependent Power Spectra}

We study the energy dependent PDS for all the `softer' variability classes to examine the HFQPO features.
While doing so, we divide $3-60$ keV energy band into different energy intervals and search for HFQPO features following the same method as discussed in section \ref{s:sps}. We find that for $\gamma$ class observation, the HFQPOs are present only in $6-25$ keV energy range with higher rms amplitude ($\sim 3.64\%$) and higher significance ($\sigma \sim 12$) at $1 \sigma$ unit compared to that in $3-60$ keV energy band (see Table \ref{table:PDS_parameters} and \ref{table:noHFQPO}). What is more is that the PDS corresponding to $3-6$ keV and $25-60$ keV energy bands are found to be featureless (see Fig. \ref{fig:PDS_zoom}) as the $rms\%$ corresponding to the most significant peak are obtained as $\sim 0.92$ and $\sim 0.74$, respectively. In addition, we model the PDS in $3-6$ keV and $25-60$ keV energy bands using \texttt{Lorentzian} function with centroid fixed at $\sim 72.32$ Hz for $\gamma$ class \citep[see][]{Belloni-etal2001, Sreehari-etal2020}. The highest values of the significance of the fitted \texttt{Lorentzian} are found to be $1.63\sigma$ and $0.7\sigma$ in $3-6$ keV and $25-60$ keV energy bands, respectively. These results clearly indicate the non-detection of HFQPO feature as the percentage rms are insignificant. We follow same methodology for all the observations under consideration and find that the prominent HFQPO features are present only in $6-25$ keV energy range (see Fig. \ref{fig:PDS_zoom} and Table \ref{table:noHFQPO}). In Fig. \ref{fig:PDS_zoom}, HFQPO feature is shown in each panel along with the non-detection cases at the insets.  Moreover, we carry out the wide-band spectral analysis to understand the emission mechanisms that possibly manifest the HFQPO signatures (see section \ref{s:spec}).

\subsection{Dynamical Power Spectra}
\label{s:dps}

The power spectra in different variability classes reveal that HFQPOs are not persistent but rather sporadic in nature. Therefore, we examine the dynamic nature of the power spectrum, where we search for HFQPO in each segment of duration $32$ s of the entire light curve of $1$ ms resolution. The Leahy power spectrum \citep{Leahy-etal1983} for each segment of the light curve is computed and plotted as a vertical slice corresponding to each time bin using the \texttt{stingray} package\footnote{\url{https://pypi.org/project/stingray/}} \citep{Huppenkothen-etal2019}. The frequency bin size is chosen as $2$ Hz. We use \texttt{bicubic} interpolation  \citep{Huppenkothen-etal2019} to improve clarity and smoothen the dynamic power spectrum. The power corresponding to each frequency is color coded such that yellow indicates minimum power and red indicates maximum power as shown in the colorbar of Fig. \ref{fig:Dyn_PDS}.

In the left side of Fig. \ref{fig:Dyn_PDS}, we present the light curve (top panel) of Orbit 10411, generated by combining {\it LAXPC10} and {\it LAXPC20} light curves along with the corresponding dynamic power spectrum (bottom panel). The light curve corresponds to the 
$\omega$ class variability (see Fig. \ref{fig:lcurvCCD}). It is observed that the power at frequencies around $70$ Hz is significant during `non-dips' (high counts) and is insignificant during the `dips' (low counts). The percentage rms amplitude of the HFQPO during high counts and low counts are obtained as $2.38 \pm 0.29$ and $0.67 \pm 0.18$ for $\omega$ class observation, respectively. This clearly suggests that the HFQPO of frequencies around $70$ Hz are generated when the count rate is high. We also observe similar behavior during $\gamma$ class observation. From the right side of Fig. \ref{fig:Dyn_PDS}, it is evident that the source emits at high count rate of about $9000$ cts/s throughout the observation. As a result, HFQPO is seen to be present in almost every $32$ s interval, although its power amplitude reduces during the narrow dips ($500-700$ s) present in the light curve. Overall, by analyzing the dynamic PDS, we infer that within the `softer' variability classes ($\kappa$, $\omega$ and $\gamma$), high count rate (non-dips) seems to be associated with the generation of HFQPO features as illustrated in Appendix A.

%%%%%%%%%%%%%%%%%%%%%%%%%%%% Figure-7 %%%%%%%%%%%%%%%%%%%%%%%%%%%%%
\begin{figure*}
	\begin{center}
		\includegraphics[width=0.49\textwidth]{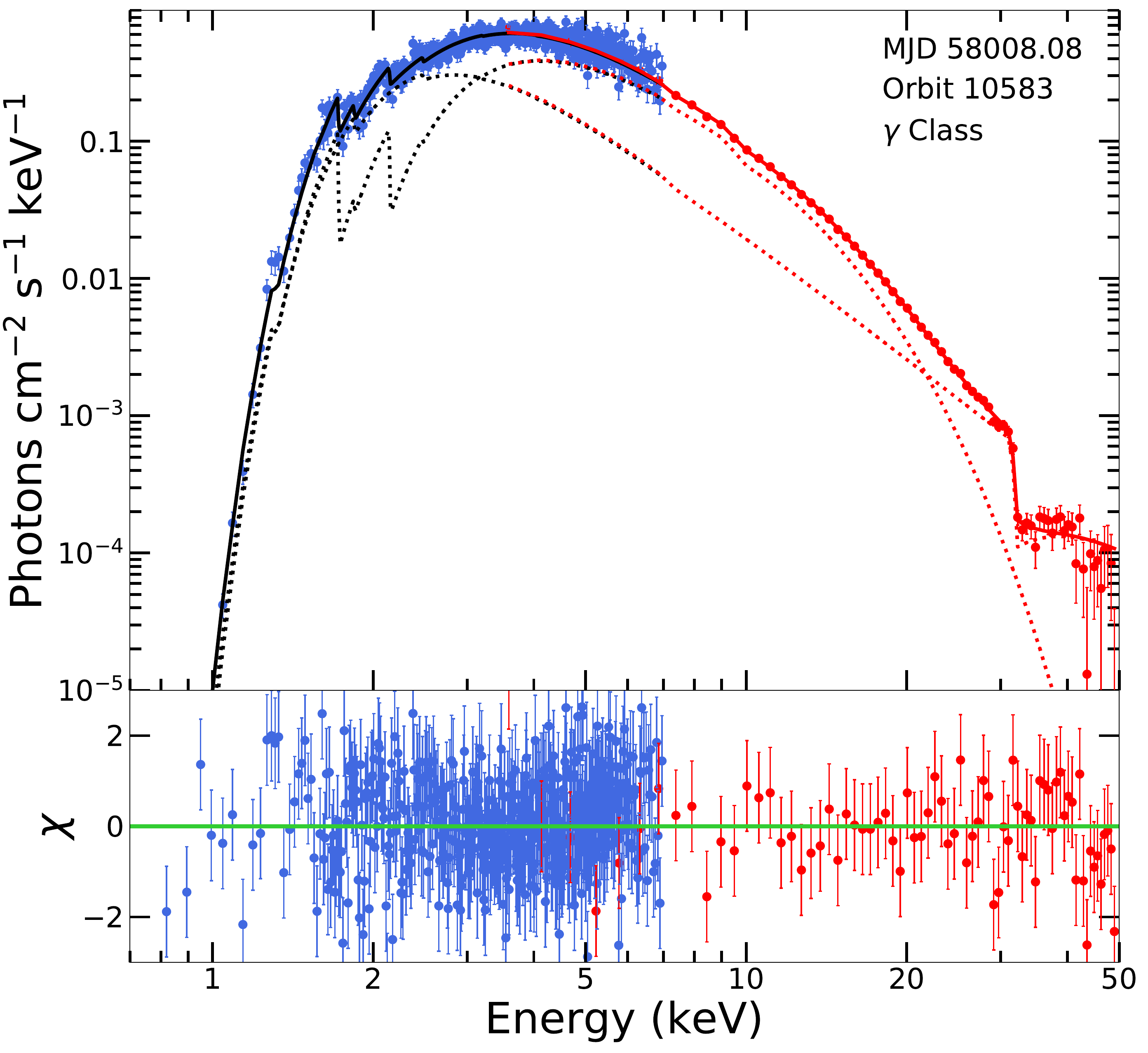}
		\includegraphics[width=0.49\textwidth]{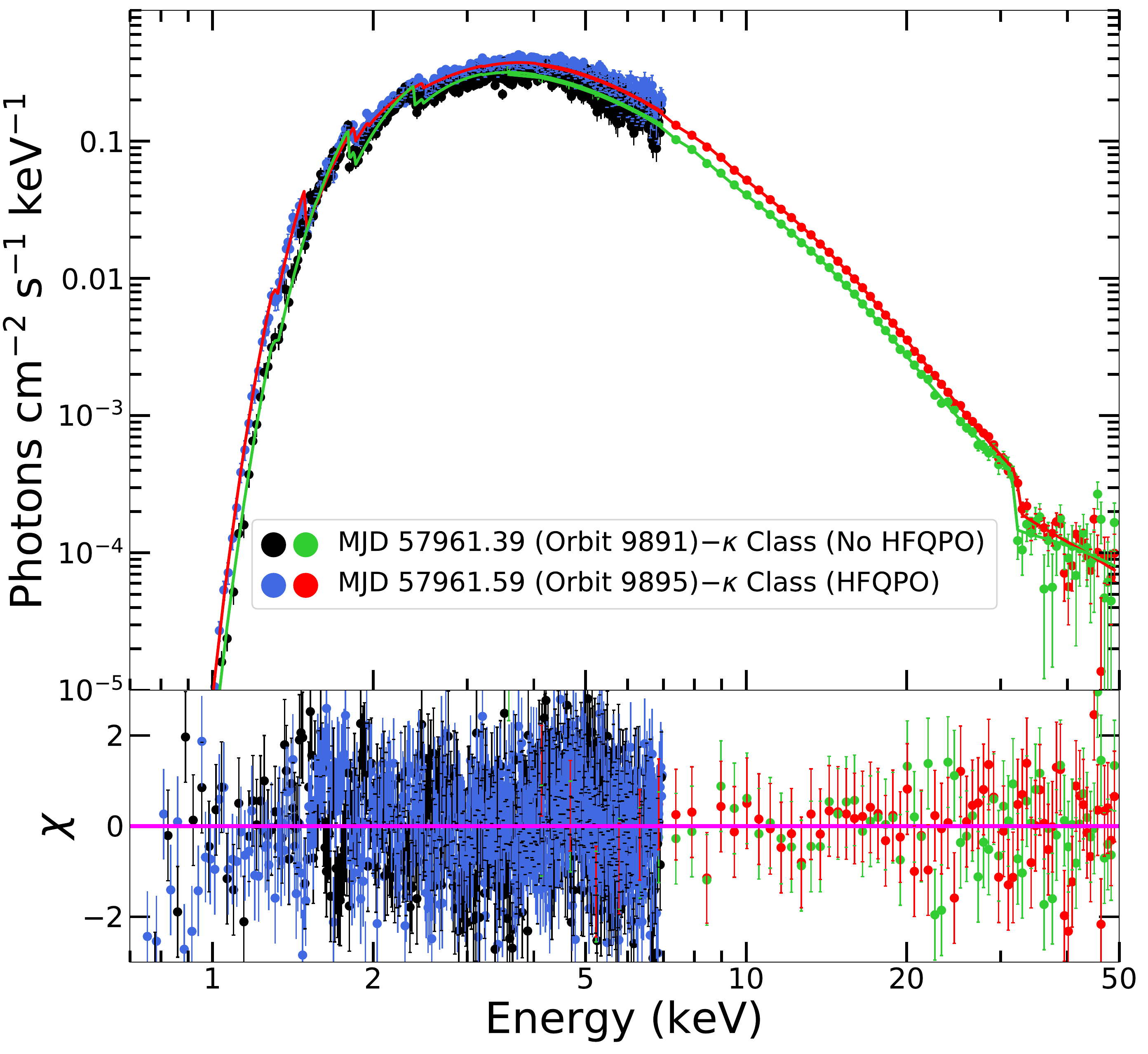}
	\end{center}
	\caption{Unfolded wide-band ($0.7-50$ keV) energy spectra of the source GRS $1915+105$. The spectra are modelled with \texttt{Tbabs}$\times$(\texttt{smedge}$\times$\texttt{nthComp} + \texttt{powerlaw})$\times$\texttt{constant}. The spectra of $\gamma$ class observation on MJD 58008.08 (Orbit 10583) is presented in the left. Two best fitted spectra of $\kappa$ class variabilities  without (MJD 57961.39; Orbit 9891) and with (MJD 57961.59; Orbit 9895) HFQPO are depicted in the right. The bottom panels of each plot show the residuals in units of $\sigma$. See text for details.}
	\label{fig:spectrum}
\end{figure*}
%%%%%%%%%%%%%%%%%%%%%%%%%%%% Figure-7 %%%%%%%%%%%%%%%%%%%%%%%%%%%%%

%%%%%%%%%%%%%%%%%%%%%%%%%%%%%%%% Table-3%%%%%%%%%%%%%%%%%%%%%%%%%%%%%%%%%%
\begin{table*}
	\caption{Estimated percentage rms amplitudes of HFQPOs in different energy bands. All the errors are computed with 68\% confidence level. See text for more details.}
	\resizebox{0.8\textwidth}{!}{% 
		\begin{tabular}{l c c c c c c c}
			\hline\hline
			\multirow{2}{*}{} & \multicolumn{3}{|c|}{HFQPO rms amplitude (\%)} & \multicolumn{1}{|c|}{Non-detection of HFQPO} & \multicolumn{1}{|c|}{Detection of HFQPO} & \multicolumn{1}{|c|}{}\\
			\cline{2-4}
			MJD (Orbit) & energy band& energy band& energy band& rms\% & HFQPO rms\%& Class& \\
			& ($3-6$ keV)& ($6-25$ keV)& ($25-60$ keV)& ($3-60$ keV)& ($3-60$ keV)& & \\	
			\hline
			57451.89 (2351)$^{\dag}$ & $-$& $-$& $-$& $0.54\pm0.23$& $-$&$\theta$&\\
			57452.82 (2365)$^{\dag}$ & $-$& $-$& $-$& $0.38\pm0.11$& $-$&$\chi$& \\
			57453.35 (2373)$^{\dag}$ & $-$& $-$& $-$& $0.44\pm0.16$& $-$&$\theta$& \\
			57504.02 (3124) & $0.39\pm0.11$& $2.33\pm0.25$& $0.81\pm0.32$& $-$& $1.48\pm0.36$ &$\omega$& \\
			57552.56 (3841)$^{\dag}$ & $-$& $-$& $-$& $0.61\pm0.23$& $-$&$\delta$& \\
			57553.88 (3860) & $0.79\pm0.28$ & $2.12\pm0.22$& $0.88\pm0.25$ & $-$& $1.54\pm0.48$&$\delta$& \\
			57689.10 (5862)$^{\dag}$ & $-$ & $-$& $-$ & $0.16\pm0.09$ & $-$ & $\beta$& \\
			57705.22 (6102)$^{\dag}$ & $-$& $-$& $-$& $0.86\pm0.15$& $-$&$\delta$& \\
			57891.88 (8863)$^{\dag}$ & $-$& $-$& $-$& $0.23\pm0.12$& $-$&$\rho \textprime$& \\
			57892.74 (8876)$^{\dag}$ & $-$& $-$& $-$& $0.77\pm0.26$& $-$&$\rho$& \\
			57943.69 (9629)$^{\dag}$ & $-$& $-$& $-$& $0.94\pm0.15$& $-$&$\kappa$& \\
			57943.69 (9633) & $0.38\pm0.14$ & $3.65\pm0.31$ & $0.56\pm0.12$& $-$& $2.14\pm0.50$ &$\kappa$& \\
			57946.10 (9666) & $0.75\pm0.18$ & $3.34\pm0.48$ & $0.62\pm0.21$ & $-$& $2.55\pm0.71$ &$\kappa$& \\
			57946.34 (9670) & $0.83\pm0.22$ & $4.29\pm0.54$ & $0.71\pm0.13$ & $-$& $2.49\pm0.37$&$\kappa$& \\
			57961.39 (9891)$^{\dag}$ & $-$& $-$& $-$& $0.82\pm0.24$& $-$&$\kappa$& \\
			57961.39 (9894) & $0.79\pm0.15$ & $2.86\pm0.29$ & $0.77\pm0.11$ & $-$& $2.28\pm0.63$ &$\kappa$& \\
			57961.59 (9895) & $0.19\pm0.07$ & $3.12\pm0.63$ & $0.84\pm0.33$ & $-$& $1.98\pm0.41$ &$\kappa$& \\
			57995.30 (10394) & $0.72\pm0.21$ & $3.30\pm0.55$& $0.67\pm0.12$ & $-$& $2.25\pm0.31$ &$\omega$& \\
			57996.46 (10411) & $0.84\pm0.24$ & $4.01\pm0.37$ & $0.81\pm0.15$ &$-$ & $2.66\pm0.29$&$\omega$& \\
			58007.80 (10579) & $0.68\pm0.12$ & $3.21\pm0.42$ & $0.89\pm0.19$ & $-$& $1.86\pm0.37$ &$\omega$& \\
			58008.08 (10583) & $0.92\pm0.16$ & $3.64\pm0.24$ & $0.74\pm0.21$ & $-$& $2.46\pm0.19$ &$\gamma$& \\
			58046.36 (11154)$^{\dag}$ & $-$& $-$& $-$& $0.42\pm0.14$& $-$&$\delta$& \\
			58209.13 (13559)$^{\dag}$ & $-$& $-$& $-$& $0.76\pm0.13$& $-$&$\chi$& \\
			58565.82 (18839)$^{\dag}$ & $-$& $-$& $-$& $0.69\pm0.11$& $-$&$\chi$& \\
			\hline
			\hline
		\end{tabular}%
	}
   \label{table:noHFQPO}
	\begin{list}{}{}
		 \item $^{\dagger}$ Non-detection of HFQPO.
	\end{list}
\end{table*}
%%%%%%%%%%%%%%%%%%%%%%%%%%%%%%%%%%%%%%%%%%%%%%%%%%%%%%%%%%%%%%%%%%%%%%%%%

%%%%%%%%%%%%%%%%%%%%%%%%%%%%%%%% Table-4 %%%%%%%%%%%%%%%%%%%%%%%%%%%%%%%%
\begin{table*}
	\caption{Best fitted parameters of the wide-band ($0.7-50$ keV) energy spectra of GRS 1915$+$105 with the model \texttt{Tbabs$\times$(\texttt{smedge}$\times$nthComp $+$ powerlaw)$\times$constant} for different observations as mentioned in Table \ref{table:Obs_details}. In the table, $kT_{e}$ is the electron temperature in keV, $\Gamma_{\rm nth}$ is the \texttt{nthComp} photon index, $norm_{\rm nth}$ is the \texttt{nthComp} normalization. $\Gamma_{\rm PL}$ and $norm_{\rm PL}$ are the \texttt{powerlaw} photon index and normalization, respectively. $F_{\rm nth}$ and $F_{\rm PL}$ are the flux in erg cm$^{-2}$ s$^{-1}$ associated with \texttt{nthComp} and \texttt{powerlaw} components, respectively. $F_{\rm bol}$, $L_{\rm bol}$, $\tau$ and y-par are the bolometric flux, bolometric luminosity in units of  $\%L_{\rm Edd}$, optical depth and Compton y-parameter, respectively. All the errors are computed with $90$ per cent confidence level. See text for details.}
	
	\resizebox{1.0\textwidth}{!}{% 
		\begin{tabular}{l @{\hspace{0.2cm}} c @{\hspace{0.22cm}} c @{\hspace{0.2cm}} c @{\hspace{0.2cm}} c @{\hspace{0.2cm}} c @{\hspace{0.2cm}} c @{\hspace{0.1cm}} c @{\hspace{0.2cm}} c @{\hspace{0.25cm}} c @{\hspace{0.25cm}} c @{\hspace{0.25cm}} c @{\hspace{0.25cm}} c @{\hspace{0.4cm}}c}
			\hline
			\hline
			& \multicolumn{5}{|c|}{Model fitted parameters} &  & \multicolumn{6}{|c|}{Estimated parameters} & \\
			
			\cline{2-6}
			\cline{8-13}
			& & & & & & & & & & & & \\
			MJD (Orbit) & $kT_{e}$ & $\Gamma_{\rm nth}$ & $norm_{\rm nth}$ & $\Gamma_{\rm PL}$ & $norm_{\rm PL}$ & $\chi_{\rm red}^2$ & $F_{\rm nth}$ & $F_{\rm PL}$ & $F_{\rm bol}$ & $L_{\rm bol}$& $\tau$ & y-par & Class\\
			
			& (keV)& & & & & & ($0.7-50$ keV)& ($0.7-50$ keV)& ($0.3-100$ keV) $^\boxplus$& ( {$\%L_{\rm Edd}$})& & \\
			
			& & & & & & & ($\tiny{10^{-8}}$ erg ${\rm cm}^{-2} {\rm s}^{-1}$)& ($\tiny{10^{-8}}$ erg ${\rm cm}^{-2} {\rm s}^{-1}$)& ($\tiny{10^{-8}}$ erg ${\rm cm}^{-2} {\rm s}^{-1}$)& & & \\
			
			\hline
			 {57451.89 (2351)}$^{\dag}$ &  {$3.66_{-0.12}^{+0.11}$}&  {$2.89_{-0.05}^{+0.08}$}&  {$34_{-2}^{+3}$}&  {$2.85_{-0.07}^{+0.07}$}&  {$42_{-3}^{+4}$}&  {1.14}&  {2.51} &  {2.84} &  {5.38} &  {30} &  {6} &  {0.83$\pm$0.06} &  {$\theta$} \\
			
			 {57452.82 (2365)}$^{\dag, \mathparagraph}$ &  {$23.92_{-4.72}^{+7.54}$}&  {$2.57_{-0.03}^{+0.03}$}&  {$18_{-1}^{+1}$}& $-$ & $-$ &  {1.06}&  {2.45} & $-$ &  {3.03}&  {17} &  {2} &  {0.63$\pm$0.08} &  {$\chi$} \\
			
			 {57453.35 (2373)}$^{\dag}$ &  {$3.10_{-0.07}^{+0.08}$} &  {$2.47_{-0.02}^{+0.03}$}&  {$29_{-2}^{+4}$}&  {$2.59_{-0.03}^{+0.03}$}&  {$23_{-2}^{+2}$}&  {1.08}&  {2.85} &  {3.14} &  {6.05} &  {34} &  {7} &  {1.29$\pm$0.04} &  {$\theta$} \\
			
			 {57504.02 (3124)} &  {$2.37_{-0.11}^{+0.12}$}&  {$1.97_{-0.05}^{+0.06}$}&  {$16_{-2}^{+3}$}&  {$2.65_{-0.08}^{+0.09}$}&  {$4_{-1}^{+2}$}&  {1.18}&  {4.46} &  {0.44} &  {5.06} &  {28} &  {12} &  {2.47$\pm$0.21} &  {$\omega$} \\
			
			57552.56 (3841)$^{\dag}$ & $1.94_{-0.05}^{+0.04}$& $1.83_{ {-0.08}}^{+0.06}$& $ {16_{-4}^{+3}}$& $3.27_{ {-0.06}}^{ {+0.05}}$& $ {34_{-7}^{+6}}$& 1.19& 3.52 & 1.71 & 5.24 &  {29} &  {14} & 3.12$\pm$ {0.35} & $\delta$ \\
			
			57553.88 (3860) & $2.53_{ {-0.11}}^{ {+0.10}}$& $2.08_{ {-0.11}}^{ {+0.09}}$& $ {19_{-5}^{+5}}$& $3.18_{ {-0.04}}^{ {+0.05}}$& $ {32_{-3}^{+4}}$& 1.12& 3.57 & 1.15 & 4.73 &  {27} &  {10} & 2.11$\pm$ {0.31}& $\delta$ \\
			
			 {57689.10 (5862)}$^{\dag}$ &  {$1.82_{-0.09}^{+0.11}$}&  {$2.13_{-0.06}^{+0.06}$}&  {$12_{-3}^{+4}$}&  {$2.83_{-0.07}^{+0.07}$} &  {$20_{-5}^{+4}$} &  {1.11}&  {1.64} &  {1.91} &  {3.85}&  {22} &  {12} &  {2.06$\pm$0.16} &  {$\beta$} \\
			
			57705.22 (6102)$^{\dag}$ & $1.93_{ {-0.04}}^{ {+0.06}}$& $1.90_{ {-0.05}}^{ {+0.05}}$& $ {15_{-2}^{+4}}$& $3.21_{ {-0.03}}^{ {+0.04}}$& $ {24_{-2}^{+3}}$& 1.08& 2.37 & 1.67 & 4.26 &  {24} &  {14} & 2.80$\pm$ {0.21}& $\delta$ \\
			
			57891.88 (8863)$^{\dag}$ & 10$^{*}$ & $2.30_{ {-0.02}}^{ {+0.04}}$& $ {9_{-1}^{+2}}$& $-$& $-$& 1.22& 1.03 & $-$& 1.05 &  {6} & $-$& $-$& $\rho \textprime$ \\
			
			57892.74 (8876)$^{\dag}$ & 10$^{*}$ & $2.86_{ {-0.07}}^{ {+0.04}}$& $ {13_{-2}^{+3}}$& $-$& $-$& 1.17& 1.02 & $-$ & 1.16 &  {7} & $-$& $-$& $\rho$ \\
			
			57943.69 (9629)$^{\dag}$ & $2.38_{ {-0.06}}^{ {+0.06}}$& $1.96_{ {-0.07}}^{ {+0.08}}$& $ {6_{-1}^{+1}}$& $2.73_{ {-0.05}}^{ {+0.05}}$& $ {2_{-1}^{+1}}$& 1.02& 1.04 & 0.59 & 1.65 &  {9} &  {12}& 2.51$\pm$ {0.29}& $\kappa$ \\
			
			57943.69 (9633) & $2.36_{ {-0.05}}^{ {+0.06}}$& $2.11_{ {-0.09}}^{ {+0.07}}$& $ {7_{-2}^{+1}}$& $3.26_{ {-0.03}}^{ {+0.04}}$ & $ {15_{-1}^{+3}}$& 1.18& 1.06& 0.52 & 1.57 &  {9} &  {11}& 2.04$\pm$ {0.21}& $\kappa$ \\
			
			57946.10 (9666) & $2.39_{ {-0.08}}^{ {+0.11}}$& $2.03_{ {-0.05}}^{ {+0.06}}$& $ {7_{-1}^{+2}}$& $3.13_{ {-0.04}}^{ {+0.04}}$& $ {13_{-2}^{+3}}$& 1.03& 1.21 & 0.54 & 1.76 &  {10} &  {11} & 2.27$\pm$ {0.19}& $\kappa$ \\
			
			57946.34 (9670) & $2.50_{ {-0.06}}^{ {+0.07}}$& $2.06_{ {-0.04}}^{ {+0.04}}$& $ {7_{-1}^{+1}}$& $3.24_{ {-0.03}}^{ {+0.03}}$& $ {16_{-1}^{+1}}$& 1.17& 1.10 & 0.55 & 1.78 &  {10} &  {11} & 2.17$\pm$ {0.12}& $\kappa$ \\
			
			57961.39 (9891)$^{\dag}$ & $2.48_{ {-0.08}}^{ {+0.09}}$& $1.97_{ {-0.06}}^{ {+0.06}}$& $ {7_{-1}^{+1}}$& $2.80_{ {-0.06}}^{ {+0.05}}$& $ {6_{-2}^{+1}}$& 1.23& 1.03 & 0.84 & 2.03 &  {12} &  {11} & 2.46$\pm$ {0.21}& $\kappa$ \\
			
			57961.39 (9894) & $2.64_{ {-0.07}}^{ {+0.10}}$& $2.10_{ {-0.05}}^{ {+0.07}}$& $ {10_{-1}^{+1}}$& $3.05_{ {-0.04}}^{ {+0.05}}$& $ {14_{-1}^{+1}}$& 1.12& 1.52 & 0.83 & 2.48 &  {14} &  {10}& 2.04$\pm$ {0.19}& $\kappa$ \\
			
			57961.59 (9895) & $2.77_{ {-0.12}}^{ {+0.11}}$& $2.23_{ {-0.12}}^{ {+0.07}}$& $ {16_{-2}^{+3}}$& $3.12_{ {-0.05}}^{ {+0.05}}$& $ {18_{-1}^{+1}}$& 1.06& 1.53 & 0.98 & 2.53 &  {14} &  {9}& 1.72$\pm$ {0.25}& $\kappa$ \\
			
			57995.30 (10394) & $3.06_{ {-0.08}}^{ {+0.13}}$& $2.44_{ {-0.08}}^{ {+0.05}}$& $ {33_{-5}^{+4}}$& $3.12_{ {-0.03}}^{ {+0.04}}$& $ {26_{-2}^{+4}}$& 1.06& 1.74 & 1.37& 3.23&  {18} &  {8}& 1.33$\pm$ {0.12}& $\omega$ \\
			
			57996.46 (10411) & $3.07_{ {-0.11}}^{ {+0.11}}$& $2.36_{ {-0.05}}^{ {+0.05}}$& $ {27_{-3}^{+3}}$& $3.13_{ {-0.04}}^{ {+0.04}}$& $ {24_{-1}^{+1}}$& 1.17& 1.94 & 1.36 & 3.43&  {19} &  {8}& 1.46$\pm$ {0.08}& $\omega$ \\
			
			58007.80 (10579) & $2.67_{ {-0.11}}^{ {+0.09}}$& $2.17_{ {-0.11}}^{ {+0.06}}$& $ {21_{-4}^{+3}}$& $3.10_{ {-0.04}}^{ {+0.03}}$& $ {26_{-1}^{+2}}$& 0.98& 2.24 & 1.31 & 3.85 &  {22} &  {9} & 1.86$\pm$ {0.14}& $\omega$ \\
			
			58008.08 (10583) & $2.62_{ {-0.12}}^{ {+0.10}}$& $2.02_{ {-0.07}}^{ {+0.06}}$& $ {15_{-2}^{+2}}$& $3.00_{ {-0.06}}^{ {+0.05}}$& $ {21_{-3}^{+4}}$& 1.08& 2.49 & 1.37 & 4.16 &  {23} &  {11}& 2.27$\pm$ {0.22}& $\gamma$ \\
			
			58046.36 (11154)$^{\dag}$ & $2.03_{ {-0.06}}^{ {+0.08}}$& $1.84_{ {-0.05}}^{ {+0.05}}$& $ {15_{-2}^{+3}}$& $2.99_{ {-0.06}}^{ {+0.07}}$& $ {14_{-1}^{+2}}$& 1.25& 3.36 & 1.68 & 5.12&  {29} &  {14} & 3.05$\pm$ {0.24}& $\delta$ \\
			
			 {58209.13 (13559)}$^{\dag, \mathparagraph}$ &  {20$^{*}$} &  {$2.19_{-0.04}^{+0.04}$}&  {$8_{-1}^{+1}$}& $-$ & $-$ &  {0.93}&  {2.21} & $-$ &  {2.34} &  {13} & $-$ & $-$ &  {$\chi$} \\
			
			 {58565.82 (18839)}$^{\dag, \mathparagraph}$ &  {20$^{*}$}&  {$1.88_{-0.02}^{+0.02}$}&  {$0.71_{-0.02}^{+0.03}$}& $-$ & $-$ &  {1.03}&  {0.51} & $-$ &  {0.53} &  {3} & $-$ & $-$ &  {$\chi$} \\
			
			\hline
			\hline
			\label{table:Spectral_parameters}
		\end{tabular}%
	}
	\begin{list}{}{}
		\item [$^\boxplus$]{\it SXT} response extends from $0.3$ keV. $^\dag$ No detection of HFQPO.
		\item [$*$] Frozen at $10$  {and $20$ keV}, below this value the spectral fitting is affected yielding higher $\chi^2_{red}$.
		\item [$^{\mathparagraph}$]  {Modelled with \texttt{diskbb} and \texttt{nthComp} component. Inner disc temperature, $T_{in} \sim 0.25$ keV.}
	\end{list}
\end{table*}

\section{Spectral Analysis and results}
\label{s:spec}

For each variability class, we generate the corresponding wide-band energy spectra combining the {\it SXT} and {\it LAXPC20} data. We consider $0.7 - 7$ keV energy range for {\it SXT} spectra, whereas {\it LAXPC} spectra are extracted in the energy range of $3-50$ keV \citep[see][for details]{Sreehari-etal2019,Sreehari-etal2020}. The dead-time corrections are applied to the {\it LAXPC} spectra while extracting
with \texttt{LaxpcSoftv3.4}\footnote{\url{http://www.tifr.res.in/~astrosat\_laxpc/LaxpcSoft.html}} software \citep{Antia-etal2017}.
	 
We model the wide-band energy spectra using \texttt{XSPEC V12.10.1f} in \texttt{HEASOFT V6.26.1} to understand the radiative emission processes active around the source. While modelling the energy spectra, we consider a systematic error of 2$\%$ for both {\it SXT} and {\it LAXPC} data \citep{Antia-etal2017, Leahy-etal2019, Sreehari-etal2020}. We use the \texttt{gain fit} command in \texttt{XSPEC} to take care of the instrumental features at $1.8$ keV and $2.2$ keV in {\it SXT} spectra. While applying gain fit, we allow the offset to vary and fix the slope at $1$. The hydrogen column density (nH) is kept fixed at $6 \times 10^{22}$ atoms/cm$^2$ following \cite{Yadav-etal2016,Sreehari-etal2020}. 

To begin with, we adopt a model \texttt{Tbabs$\times$nthComp$\times$constant} to fit the wide-band energy spectrum of $\gamma$ class observation (Orbit 10583) for which the strongest HFQPO feature is seen in the PDS (bottom panel of Fig. \ref{fig:PDS_QPO}). Here, the model \texttt{Tbabs} \citep{Wilms-etal2000} takes care of the galactic absorption between the source and the observer. The model \texttt{nthComp} \citep{Zdziarski-etal1996} represents the thermally Comptonized continuum. A \texttt{constant} parameter is used to account for the offset between the spectra from two different instruments, {\it SXT} and {\it LAXPC}. The obtained fit is yielded a poor reduced $\chi^2$ ($\chi_{\rm red}^{2}=\chi^{2}/dof$) of $2095/546=3.83$ as there are large residuals left at the higher energies beyond $30$ keV. Hence, we include a \texttt{powerlaw} along with one Xenon \texttt{edge} component at $32$ keV \citep{Sreehari-etal2019} to fit the high energy part of the spectrum. The Xenon \texttt{edge} is required for all the energy spectra to account for the instrumental absorption feature at $32-35$ keV. Also, one \texttt{smedge} component is used at $\sim 9$ keV to obtain the best fit. The model \texttt{Tbabs$\times$(\texttt{smedge}$\times$nthComp $+$ powerlaw)$\times$constant} provides statistically acceptable fit with $\chi_{\rm red}^{2}=\chi^{2}/dof$ of $563/533=1.08$. In the left side of Fig. \ref{fig:spectrum}, we present the best fitted unfolded energy spectrum of $\gamma$ class observation for representation.

The best fitted model parameters of the \texttt{nthComp} component for $\gamma$ class observation (Orbit 10583) are obtained as electron temperature  $(kT_{e}) = 2.62^{+0.10}_{-0.12}$ keV, photon index  $(\Gamma_{\rm nth}) = 2.02^{+0.06}_{-0.07}$ with normalization  $(norm_{\rm nth}) = 15\pm2$. The seed photon temperature ($kT_{\rm bb}$) is kept fixed at $0.1$ keV during the fitting. The best fitted value for the \texttt{powerlaw} photon index ($\Gamma_{\rm PL}$) is obtained as $\Gamma_{\rm PL} = 3.00^{+0.05}_{-0.06}$ with \texttt{powerlaw} normalization  $norm_{\rm PL} = 21^{+4}_{-3}$. Following the same approach, we carry out the spectral modelling for all other observations of various variability classes ($i.e.$, $\theta$, $\beta$, $\delta$, $\rho$, $\kappa$, $\omega$, $\gamma$ and $\chi$) irrespective to the presence or absence of HFQPOs. The best fitted model parameters are tabulated in Table \ref{table:Spectral_parameters}. It is found that in all observations, the model \texttt{Tbabs$\times$(\texttt{smedge}$\times$nthComp $+$ powerlaw)$\times$constant} satisfactorily describes the energy spectra except for $\rho$ and $\chi$ class observations. For $\rho$ class, the acceptable fit is obtained without \texttt{powerlaw} component, whereas in $\chi$ class observation, an additional \texttt{diskbb} component is required along with \texttt{nthComp}. Note that we are unable to constrain the electron temperature for the observations of $\rho$, $\rho^\prime$ and two $\chi$ classes and hence we fix the electron temperature $kT_e = 10$ keV and $20$ keV, respectively (see Table \ref{table:Spectral_parameters}). In the right side of Fig. \ref{fig:spectrum}, we present the wide-band energy spectra of two $\kappa$ class observations on MJD 57961.39 (Orbit 9891) and MJD 57961.59 (Orbit 9895) without and with HFQPO feature, respectively. We point out that the best fitted energy spectrum of orbit 9891 (without HFQPO) has a weak \texttt{nthComp} contribution ( $norm_{\rm nth} \sim 7$), whereas in orbit 9895 (with HFQPO), the \texttt{nthComp} contribution ( $norm_{\rm nth} \sim 16$) is relatively higher. We also find relatively high electron temperatures in those observations that ascertain the detection of HFQPOs except one observation in $\theta$ class (Orbit 2351).

Further, we estimate the flux in the energy range $0.7-50$ keV associated with different model components used for the spectral fitting. While doing so, the convolution model \texttt{cflux} in \texttt{XSPEC} is used.
For $\gamma$ class observation, the fluxes associated with \texttt{nthComp} and \texttt{powerlaw} components are estimated as $2.49$ and $1.37$ in units of $10^{-8}$ erg ${\rm cm}^{-2} ~{\rm s}^{-1}$, respectively (see Table \ref{table:Spectral_parameters}). We also calculate bolometric flux ($F_{\rm bol}$) in the energy range $0.3-100$ keV using the \texttt{cflux} model. Considering the mass ($M_{\rm BH}$) and the distance ($d$) of the source as $M_{\rm BH} =12.4 M_\odot$ and $d=8.6$ kpc \citep{Reid-etal2014}, we calculate the bolometric luminosity in units of Eddington luminosity ($L_{\rm Edd}$)\footnote{Eddington luminosity $L_{\rm Edd}=1.26\times10^{38} (M_{\rm BH}/M_\odot)$ erg $\rm s^{-1}$ for a compact object of mass $M_{\rm BH}$ \citep{Frank-etal2002}.} as $L_{\rm bol}=F_{\rm bol}\times4\pi d^{2}$. The luminosity is found to vary in the range of $3-34\%$ $L_{\rm Edd}$. The calculated flux values and bolometric luminosities for all the observations are given in Table \ref{table:Spectral_parameters}.

In order to understand the nature of the Comptonizing medium in the vicinity of the source, we calculate the optical depth ($\tau$) of the medium. Following \cite{Zdziarski-etal1996, Chatterjee-etal2021}, the relation among the optical depth ($\tau$), \texttt{nthComp} spectral index ($\alpha=\Gamma_{\rm nth}-1$), and electron temperature ($kT_{e}$) is given by,
$$
\alpha=\left[\frac{9}{4}+\frac{1}{\left(\nicefrac{kT_{e}}{m_{e}c^{2}}\right)\tau\left(1+\nicefrac{\tau}{3}\right)}\right]^{\unitfrac{1}{2}}-\frac{3}{2},
\eqno(1)
$$
where $m_e$ is the electron mass, and $c$ refers to the speed of light. Using equation (1), the optical depth ($\tau$) is calculated and found in the range
$2 \lesssim \tau \lesssim 14$ for all the observations. This implies that an optically thick corona is present as a Comptonizing medium in the vicinity of the source. Moreover, we calculate the Compton y-parameter, which measures the degree of Compton up-scattering of soft photons in the underlying accretion flow by the Comptonizing medium. Following \cite{Agrawal-etal2018, Chatterjee-etal2021}, the Compton y-parameter ($=4kT_{e}\tau^{2}/m_{e}c^{2}$) in the optically thick medium is found to be in the range of $0.63\pm0.08 ~-~ 3.12\pm0.35$. In Table \ref{table:Spectral_parameters}, we tabulate the obtained optical depth ($\tau$) and Compton y-parameter values for all the observations. 

\section{Spectro-temporal correlation}

%%%%%%%%%%%%%%%%%%%%%%%%%%%% Figure 8 %%%%%%%%%%%%%%%%%%%%%%%%%%%%%%%
\begin{figure}
	\begin{center}
		\includegraphics[width=\columnwidth]{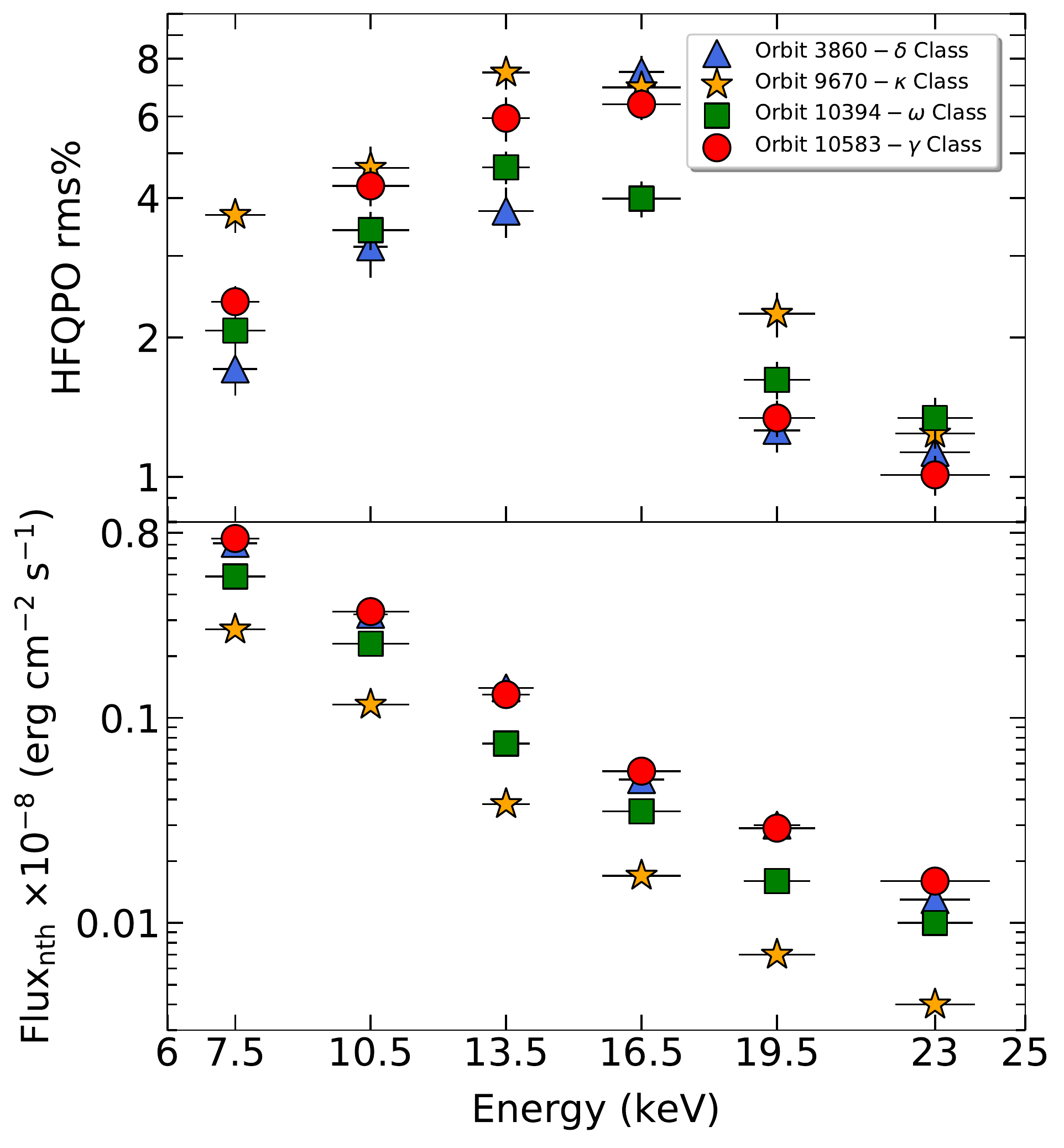}
	\end{center}

	\caption{{\it Top}: Variation of percentage rms amplitudes ($rms\%$) of HFQPOs with energy for $\delta$, $\kappa$, $\omega$ and $\gamma$ variability classes. {\it Bottom}: Variation of \texttt{nthComp} flux with energy corresponding to the variability classes as marked in the inset of the top panel. See text for details.  
	}
	\label{fig:rms_energy}
\end{figure}
%%%%%%%%%%%%%%%%%%%%%%%%%%%%%%%%%%%%%%%%%%%%%%%%%%%%%%%%%%%%%%%%%%%%%

In this section, we examine the spectro-temporal correlation of the observed properties in different variability classes of GRS 1915+105. While doing so, we consider four variability classes, namely $\delta$, $\kappa$, $\omega$ and $\gamma$, and study the variation of the percentage rms amplitude ($rms\%$) of HFQPO features as function of energy shown in the top panel of Fig. \ref{fig:rms_energy}. The results corresponding to $\delta$, $\kappa$, $\omega$ and $\gamma$ classes are denoted by the filled triangles (blue), asterisks (yellow), squares (green) and circles (red), respectively. We find that HFQPO $rms\%$ increases with energy up to $\sim 17$ keV and then sharply decreases. Further, to correlate the evolution of $rms\%$ with the Comptonized flux ($i.e.$, \texttt{nthComp}), we present the variation of the estimated \texttt{nthComp} flux in the respective energy bands as shown in the bottom panel of Fig. \ref{fig:rms_energy}. We notice that \texttt{nthComp} flux decreases with energy and it becomes negligible beyond $\sim 25$ keV ($\sim 2\%$ of the total \texttt{nthComp} flux in $3-60$ keV; see also Fig. \ref{fig:spectrum}).

%%%%%%%%%%%%%%%%%%%%%%%%%%%% Figure 9 %%%%%%%%%%%%%%%%%%%%%%%%%%
\begin{figure}
	\begin{center}
		\includegraphics[width=\columnwidth]{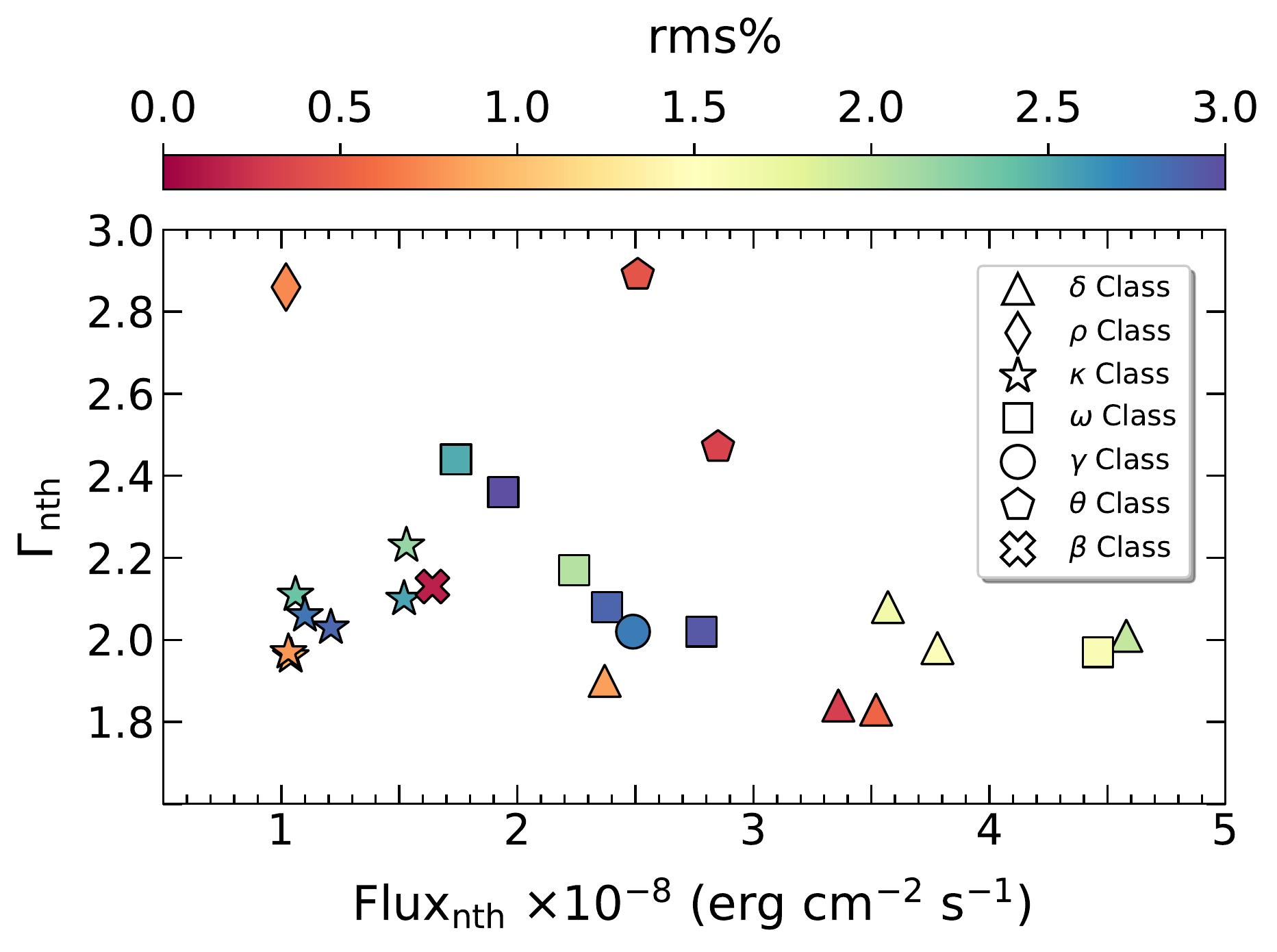}
	\end{center}
	\caption{Variation of photon index ($\Gamma_{\rm nth}$) as a function of Comptonized flux associated with \texttt{nthComp} (see Table \ref{table:Spectral_parameters}) for $\delta$, $\rho$, $\kappa$, $\omega$, $\gamma$, $\theta$ and $\beta$ variability classes. The color codes denote the rms amplitudes ($rms\%$) corresponding to the detection and non-detection of HFQPOs. Obtained results for different variability classes are presented using different symbols which are marked in the inset. See text for details.}
	\label{fig:flux_gamma}
\end{figure}
%%%%%%%%%%%%%%%%%%%%%%%%%%%%%%%%%%%%%%%%%%%%%%%%%%%%%%%%%%%%%%%%%
%

In Fig. \ref{fig:flux_gamma}, we present the variation of photon index ($\Gamma_{\rm nth}$) with the \texttt{nthComp} flux ($F_{\rm nth}$) for observations in $\delta$, $\rho$, $\kappa$, $\omega$, $\gamma$, $\beta$, and $\theta$ variability classes. Here, the variation of the percentage rms amplitude ($rms\%$) of the detected HFQPOs as well as the $rms\%$
of the non-detection of HFQPOs (see Table \ref{table:noHFQPO}) are shown using color code. We find that $rms\%$ of HFQPOs lies in the range $\gtrsim$ 1.5\% and it drops below $1\%$ when HFQPOs are not seen. Further, we notice that in $\omega$ class observations, $\Gamma_{\rm nth}$ decreases with the increase of \texttt{nthComp} flux in the range $1.8 \times 10^{-8}\lesssim {\rm Flux}_{\rm nth} ~({\rm erg ~cm}^{-2}~ {\rm s}^{-1}) \lesssim 4.5 \times 10^{-8}$. In $\kappa$ variability classes, marginal variations are observed in the \texttt{nthComp} flux unlike $\omega$ class observation although photon indices remain $\Gamma_{\rm nth} \gtrsim 2$ when HFQPOs are present. For $\delta$ variability classes, we find $3.5 \times 10^{-8} \lesssim {\rm Flux}_{\rm nth} ~({\rm erg ~cm}^{-2}~ {\rm s}^{-1}) \lesssim 4.58 \times 10^{-8} $ and $\Gamma_{\rm nth} \gtrsim 1.9$ for HFQPOs. In addition, we point out that HFQPOs are not seen for $\theta$, $\beta$ and $\rho$ class observations although they belong to the `softer' variability classes.

\section{Discussion}
\label{s:dis}

In this work, we carry out a comprehensive timing and spectral analyses of the BH-XRB source GRS 1915+105 using GT, AO and TOO data of entire {\it AstroSat} observations ($2016-2021$) in wide frequency ($0.01 - 500$ Hz) and energy ($0.7 - 60$ keV) bands. The variation of the hardness ratios in the CCDs and the nature of the light curves confirm the presence of seven `softer' variability classes, namely $\theta$, $\beta$, $\delta$, $\rho$, $\kappa$, $\omega$, and $\gamma$ along with one additional variant of $\rho$ class ($\rho \textprime$) of the source \citep{Athulya-etal2021}. The {\it LAXPC} count rates are found to vary in the range of $1328-10997$ cts/s, and the hardness ratios are seen to vary as $0.61 \lesssim {\rm HR1} \lesssim 0.90$ and $0.02 \lesssim {\rm HR2} \lesssim 0.11$. We observe similar variabilities in the low energy ($0.7-7$ keV) {\it SXT} light curves as well. 

We examine the origin of the HFQPO features and find its presence only in four `softer' variability classes ($\delta$, $\omega$, $\kappa$, $\gamma$). The centroid frequency and the percentage rms amplitude of the HFQPOs are found to vary in the range of $68.14-72.32$ Hz and  $1.48-2.66$ per cent, respectively (see Table \ref{table:PDS_parameters} and \ref{table:noHFQPO}). The present findings are consistent with the earlier results reported using {\it RXTE} observations \citep{Belloni-etal2013} as well as {\it AstroSat} observations \citep{Belloni-etal2019,Sreehari-etal2020}.
It may be noted that the presence of additional peaks in the frequency range $27-41$ Hz along with HFQPOs at $67-69$ Hz was also observed \citep{Belloni-etal2001, Strohmayer2001b} in GRS 1915+105. However, we do not find any signature of additional peak in our analysis. Meanwhile, 
\cite{Morgan-etal1997,Strohmayer2001b} noticed the evolution of
frequency from $\sim 67$ Hz to $\sim 69$ Hz in $\gamma$ class which is found to evolve further to $72.32^{+0.23}_{-0.21}$ Hz as seen in our observations (see Table \ref{table:PDS_parameters} and \ref{table:noHFQPO}). This perhaps indicates an unique characteristics of the evolution of HFQPO in a given variability class that requires further investigation.

Next, we investigate the energy dependent PDS to ascertain the photons that are responsible for the origin of the HFQPOs (see Fig. \ref{fig:PDS_zoom}).  
We observe that HFQPO features are present in $6-25$ keV energy band in four variability classes (see Fig. \ref{fig:PDS_zoom} and Table \ref{table:noHFQPO}). However, we do not find HFQPO signature in $\theta$, $\beta$, $\rho$ and $\chi$ variability classes (see Fig. \ref{fig:PDS_noQPO} and Table \ref{table:noHFQPO}). We find that the significance and percentage rms amplitudes of the HFQPOs are higher in $6-25$ keV energy band compared to $3-60$ keV energy band (see Table \ref{table:noHFQPO}). Further, we study the variation of percentage rms amplitude and \texttt {nthComp} flux as function of energy, shown in Fig. \ref{fig:rms_energy}. These findings are in agreement with the previous studies which were carried out considering energy bands of $2-13$ keV and $13 - 30$ keV only \citep{Belloni-etal2001}. It is also found that the \texttt{nthComp} flux gradually decreases with energy (see Fig. \ref{fig:rms_energy}) and beyond $\sim 25$ keV, the flux contribution becomes negligible (see Fig. \ref{fig:spectrum}).

We notice that $\kappa$ and $\omega$ class variabilities exhibit different duration of `non-dips' (high counts $\sim 9600$ cts/s) and `dips' (low counts $\sim 1250$ cts/s) features. While studying the dynamic PDS, we find that the HFQPO is generally present during `non-dips' period with relatively higher rms amplitude ($2.38\pm 0.29$ and $2.35\pm 0.22$ for $\omega$ and $\kappa$ classes). We do not find the signature of HFQPO during the `dips' period as seen in Fig. \ref{fig:Dyn_PDS} (see also appendix A). Similar findings are also observed for $\gamma$ class variability where the HFQPO is persistently seen during the entire high count duration \cite[see also][]{Belloni-etal2001}.

We find that the wide-band spectra ($0.7-50$ keV) of $\theta$, $\beta$, $\delta$, $\kappa$, $\omega$ and $\gamma$ class observations are satisfactorily described by the thermal Comptonization \texttt{nthComp} along with a \texttt{powerlaw} component (see Fig. \ref{fig:spectrum} and Table \ref{table:Spectral_parameters}). On the contrary, the \texttt{nthComp} component seems to be adequate to fit the energy spectra of $\rho$ class observations. From the spectral modelling, we obtain the range of \texttt{nthComp} photon index as $1.83 \lesssim \Gamma_{\rm nth} \lesssim 2.89$, and electron temperature as $1.82 \lesssim kT_{e} \lesssim 3.66$ keV (see Table \ref{table:Spectral_parameters}). In addition, we obtain a steep \texttt{powerlaw} photon index ($\Gamma_{\rm PL}$) as $2.59-3.27$ for all the variability classes under consideration. Similar steep $\Gamma_{\rm PL}$ is also reported in the previous studies carried out for $\theta$ \citep{Belloni-etal2006} and $\delta$ \citep{Sreehari-etal2020} variability classes of the source. We estimate the optical depth ($\tau$) of the surrounding medium and obtain its value as $2 \lesssim \tau \lesssim 14$. This evidently indicates the presence of a cool and optically thick corona around the source, which presumably acts as a Comptonizing medium that reprocesses the soft seed photons. The Compton y-parameter is obtained in the range $0.63\pm0.08 ~-~ 3.12\pm0.35$ which infers that the soft photons are substantially reprocessed via Comptonization at the optically thick corona. Further, we calculate the bolometric luminosity in $0.3-100$ keV energy range for all the variability classes and find its value in the range $3-34\%$ $L_{\rm Edd}$. These findings suggest that the source possibly emits in sub-Eddington limit during the observations under consideration.
 
In Fig. \ref{fig:rms_energy}, we examine the energy dependent ($6-25$ keV) $rms\%$ of HFQPOs in $\delta$, $\kappa$, $\omega$, and $\gamma$ variability classes, and find that $rms\%$ increases ($1-8\%$) with energy up to $\sim 17$ keV and then decreases. Subsequently, we observe that the Comptonize flux (${\rm Flux}_{\rm nth}$) decreases with energy which tends to become negligible beyond $25$ keV. This possibly happens when the soft photons emitted from the disc are Comptonized by the `hot' electrons from an optically thick corona ($8 \lesssim \tau \lesssim 12$) and produce the aforementioned Comptonized continuum. In order to elucidate the spectra above $\sim 25$ keV, an additional \texttt{powerlaw} component with photon index $\Gamma_{\rm PL} \sim 3$ is required. This eventually indicates that there could be an extended corona present surrounding the central corona, which is responsible for this high energy emissions \citep{Sreehari-etal2020}. We further notice that for intermediate range of \texttt{nthComp} photon index ($1.97 \lesssim \Gamma_{\rm nth} \lesssim 2.44$) along with the large variation of \texttt{nthComp} flux ($1.06 \times 10^{-8} \lesssim {\rm Flux}_{\rm nth} ~({\rm erg ~cm}^{-2}~ {\rm s}^{-1}) \lesssim 4.46\times 10^{-8} $) generally yields HFQPO signature in $\delta$, $\kappa$, $\omega$ and $\gamma$ variability classes (see Fig. \ref{fig:flux_gamma}). With this, we argue that Comptonization process plays a viable role in exhibiting HFQPO. Overall, based on the findings presented in Fig. \ref{fig:rms_energy} and Fig. \ref{fig:flux_gamma}, we conjecture that HFQPOs in GRS 1915$+$105 are perhaps manifested due to the modulation of the `Comptonizing corona' surrounding the central source \citep{Mendez-etal2013,Aktar-etal2017,Aktar-etal2018,Dihingia-etal2019,Sreehari-etal2020}.

Meanwhile, several theoretical models are put forwarded to explain the HFQPO features occasionally observed in BH-XRBs. \cite{Morgan-etal1997} first attempted to elucidate the HFQPO features with Keplerian frequency associated with the motion of hot gas at the innermost stable circular orbit (ISCO). This model yielded high source mass as $\sim 30~M_\odot$, which is in disagreement with the dynamical mass measurement of GRS 1915+105 \citep{Greiner-etal2001,Reid-etal2014} and hence dissented \citep{Belloni-etal2013}. \cite{Nowak-etal97} interpreted the origin of $67$ Hz QPO in GRS 1915+105 as the resulting frequency of the lowest radial g-mode oscillation in the accretion disc. However, this model lacks cogency for those BH-XRBs that generally manifest $>10\%$ rms amplitude variability. Further, there were alternative attempts to address the origin of HFQPO without dwelling much on observational features \citep{Chen-Taam1995,Rezolla-etal2003,Stuchlik-etal2007}. Noticing significant hard lag in GRS 1915+105, \cite{Cui-etal1999} infer that a Comptonizing region is responsible for the HFQPOs. \cite{Remillard-etal2002} further stressed on the presence of `Compton corona' that reprocesses the disk photons and yields HFQPO features. In addition, \cite{Aktar-etal2017,Aktar-etal2018} reported that the HFQPO features of GRO J1655$-$40 at $300$ Hz and $450$ Hz perhaps resulted due to the modulations of the post-shock corona (PSC) that radiates Comptonized emissions \citep{Chakrabarti-etal1995}. Recently, \cite{Dihingia-etal2019} ascertained that the shock induced relativistic accretion solutions are potentially viable to explain the HFQPOs in well studied BH-XRB sources, namely GRS 1915$+$105 and GRO J1655$-$40. With this, we affirm that the HFQPO models proposed based on the `Comptonizing corona' fervently favor our observational findings delineated in this work. 

\section{Conclusions}

In this paper, we perform in-depth temporal and spectral analyses in the wide-band ($0.7-60$ keV) energy range of the BH-XRB source GRS 1915+105 using entire {\it AstroSat} observations ($2016-2021$) in seven `softer' variability classes, namely $\theta$, $\beta$, $\delta$, $\rho$, $\kappa$, $\omega$ and $\gamma$, and one `harder' variability class ($\chi$), respectively. The overall findings of this work are summarized below:

\begin{itemize}

	\item We find HFQPO feature in $\delta$, $\kappa$, $\omega$ and $\gamma$ classes of GRS 1915 + 105 having frequency in the range $68.14 - 72.32$ Hz. However, we did not find the signature of HFQPO features in $\theta$, $\beta$, $\rho$, and $\chi$ class observations.
	
	\item Energy dependent PDS study indicates that the emergent photons in the energy range $6-25$ keV seem to be responsible for generating the HFQPO features. Beyond this energy range, the HFQPO signature is not detected. We notice that percentage rms amplitude ($rms\%$) of HFQPOs increases ($1-8\%$) with energy up to $\sim 17$ keV and then decreases.

	\item Dynamical PDS of $\kappa$ and $\omega$ classes reveal that the `non-dips' (high count) features of the light curve are possibly linked with the generation of HFQPOs.

	\item The wide-band spectral modelling indicates that in presence of HFQPO, the thermal Comptonization components ($1.06 \times 10^{-8} \lesssim F_{\rm nth} ~({\rm erg} {\rm ~cm}^{-2} {\rm ~s}^{-1}) \lesssim  4.46 \times 10^{-8}$) having $\Gamma_{\rm nth}$ of $1.97-2.44$ dominate (up to $\sim 25$ keV) over the additional {\it powerlaw} component ($0.52 \times 10^{-8} \lesssim F_{\rm PL} ~({\rm erg} {\rm ~cm}^{-2} {\rm ~s}^{-1}) \lesssim  1.37 \times 10^{-8}$) with $\Gamma_{PL}$ $\sim 3$ (above $\sim 25$ keV).

\end{itemize}

With the above findings, we argue that the variability produced during the `softer' classes of GRS 1915+105 is possibly due to the modulation of the Comptonizing corona that manifests the HFQPO features.

%%%%%%%%%%%%%%%%%%%%%%%%%%%%%%%%%%%%%%%%%%%%%%%%%%%%%%%%%%%%%%%%%%%
\section*{Acknowledgments}

Authors thank the anonymous reviewer for valuable comments and suggestions that help to improve the clarity of the manuscript. SM, NA, and SD thank the Department of Physics, IIT Guwahati, for providing the facilities to complete this work. NA, SD, and AN acknowledge the support from ISRO sponsored project (DS\_2B-13013(2)/5/2020-Sec.2). AN thanks GH, SAG; DD, PDMSA, and Director, URSC for encouragement and continuous support to carry out this research. This publication uses the data from the {\it AstroSat} mission of the Indian Space Research Organisation (ISRO), archived at the Indian Space Science Data Centre (ISSDC). This work has used the data from the Soft X-ray Telescope (SXT) developed at TIFR, Mumbai, and the SXT-POC at TIFR is thanked for verifying and releasing the data and providing the necessary software tools. This work has also used the data from the {\it LAXPC} Instruments developed at TIFR, Mumbai, and the LAXPC-POC at TIFR is thanked for verifying and releasing the data. We also thank the {\it AstroSat} Science Support Cell hosted by IUCAA and TIFR for providing the \texttt{LAXPCSOFT} software which we used for {\it LAXPC} data analysis.

\section*{Data Availability}
Data used for this publication are currently available at the Astrobrowse (AstroSat archive) website (\url{https://astrobrowse.issdc.gov.in/astro\_archive/archive}) of the Indian Space Science Data Center (ISSDC). 

\input{ms_03.bbl}
%\bibliography{references}

\appendix 

\section{Intensity dependent power spectra}

\begin{figure}
	\begin{center}
		\includegraphics[width=\columnwidth]{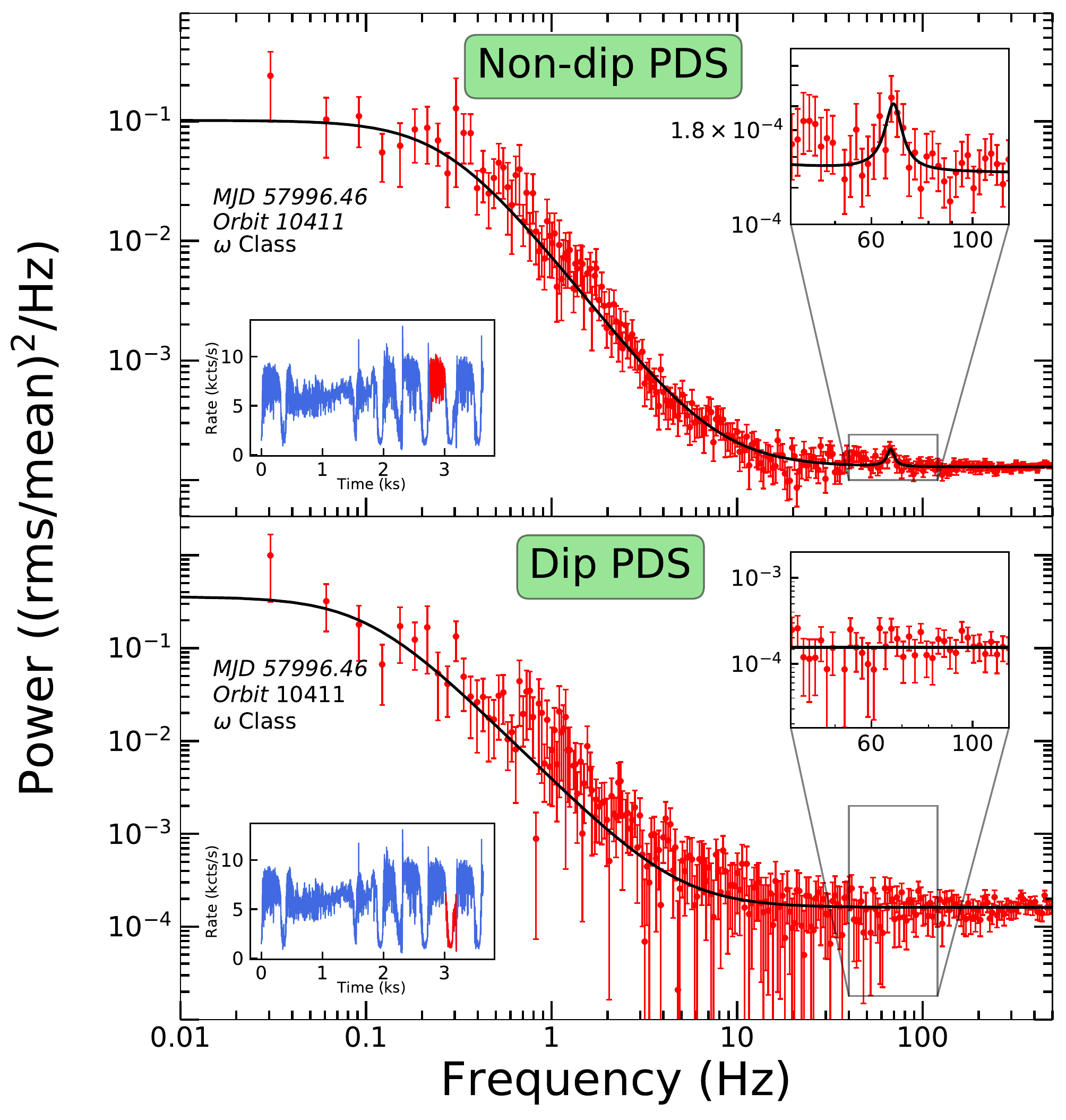}
	\end{center}
	
	\caption{{\it Top}: Power density spectrum (PDS) corresponding to one `non-dip' (high counts) duration (red segment of the light curve) of $\omega$ class observation (Orbit 10411) in a wide frequency range of $0.01-500$ Hz. {\it Bottom}: PDS corresponding to one `dip' (low counts) duration (red segment of the light curve). Both PDS are obtained for the energy band of $3-60$ keV with {\it LAXPC10} and {\it LAXPC20} combined observations. In each panel, zoomed view of the high frequency region of the PDS is depicted at the inset. 	
	}
\end{figure}
%%%%%%%%%%%%%%%%%%%%%%%%%%%%%%%%%%%%%%%%%%%%%%%%%%%%%%%%%%%%%%%%%

The power spectra are generated separately considering both `non-dips' (high counts) and `dips' (low counts) duration of the light curve (Orbit 10411, $\omega$ class). In the top panel of Figure A1, we show the power spectrum corresponding to the `non-dip' segment of the light curve shown at the bottom-left inset using red color. The HFQPO signature around $67.08^{+1.05}_{-1.35}$ Hz is distinctly visible ($rms\% = 2.38 \pm 0.29$) at the top-right inset. In the lower panel, we plot the power spectrum obtained for the `dip' segment of the light curve which is indicated in red color at the bottom-left inset. The HFQPO feature is not seen ($rms\% = 0.67\pm0.18$) as shown at the top-right inset. We further confirm that similar findings are observed during $\kappa$ class observation (Orbit 9895) as well, where $rms\%$ of the HFQPO feature for `non-dips' and `dips' duration are obtained as $2.35 \pm 0.22$ and $0.26 \pm 0.10$, respectively.

\label{lastpage}

\end{document}